\documentclass[fleqn,usenatbib]{mnras}

\usepackage[T1]{fontenc}

\DeclareRobustCommand{\VAN}[3]{#2}
\let\VANthebibliography\thebibliography
\def\thebibliography{\DeclareRobustCommand{\VAN}[3]{##3}\VANthebibliography}

\usepackage{graphicx}	%
\usepackage{amsmath}	%
\usepackage{amssymb}	%
\usepackage{bm}
\usepackage{csquotes}
\usepackage[shortlabels]{enumitem}
\usepackage{newtxtext,newtxmath}

\def\draftversion{1} %
\usepackage{amssymb,latexsym,graphicx,natbib,eufrak,times,amsmath,xspace,ifthen}

\ifthenelse{ \draftversion > 0 }{
} {
  
}

\ifthenelse{\equal{\draftversion}{1}}{
  \usepackage{xcolor}
  \newcommand{\sep}[1]{\par\begin{color}[rgb]{0,0.4,0}\begin{center}\hrule\end{center}\end{color}\par} %
  \newcommand{\todo}[1]{\begin{color}{red}\ \ifthenelse{\equal{#1}{}} {$\bullet\bullet\bullet$} {$\bullet$\ #1 $\bullet$}\end{color}} %
  \newcommand{\idea}[1]{\begin{color}[rgb]{0,0.4,0}\textit{#1}\end{color}} %
  \newcommand{\sk}[1]{\begin{color}[rgb]{0.6,0,0.6}#1\end{color}} %
  \newcommand{\toc}{\par\begin{color}[rgb]{0.6,0,0.6}\begin{center}\hrule\vspace{0.5mm}\begingroup\small\let\cleardoublepage\relax\let\clearpage\relax\mytoc\endgroup\vspace{0.5mm}\hrule\end{center}\end{color}\par} %

  }{
  \newsavebox{\trashcan}

  \newcommand{\sep}[1]{}
  \newcommand{\todo}[1]{}
  \newcommand{\idea}[1]{}
  \newcommand{\sk}[1]{}
  \newcommand{\toc}{}

  }
 \makeatletter\newcommand\mytoc{\@starttoc{toc}}\makeatother %
\long\def\symbolfootnote[#1]#2{\begingroup%
\def\thefootnote{\fnsymbol{footnote}}\footnote[#1]{#2}\endgroup} 

\newcommand{\bb}[1]{\ifmmode \mbox{\boldmath $ #1$} \else  \mbox{\boldmath $#1$} \fi}

\newcommand{\U}[1]{\ensuremath{\mathrm{~#1}}}     %

\newcommand{\Myr}{\U{Myr}\xspace}          
\newcommand{\Gyr}{\U{Gyr}\xspace}          
\newcommand{\pc}{\U{pc}\xspace}

\newcommand{\Msun}{\U{M}_{\odot}}

\newcommand{\cc}{\U{cm^{-3}}}
\newcommand{\K}{\U{K}}
    
\newcommand{\Msunpcs}{\U{M_{\odot}\ pc^{-2}}}    
\newcommand{\kms}{\U{km\ s^{-1}}}

\newcommand{\mach}{\ensuremath{\mathcal{M}}}      %
\newcommand{\tff}{\ensuremath{t_\mathrm{ff}}}     %
\newcommand{\tturb}{\ensuremath{t_\mathrm{turb}}}     %
\newcommand{\tdep}{\ensuremath{\tau_\mathrm{dep}}\xspace}        %

\newcommand{\ramses}{{\small RAMSES}\xspace}

\newcommand{\starburst}{{\small STARBURST99}\xspace}
\newcommand{\music}{{\small MUSIC}\xspace}

\newcommand{\vintergatan}{{\small VINTERGATAN}\xspace}
\newcommand{\agora}{{\small AGORA}\xspace}

\newcommand{\eobs}{$\epsilon_{\rm obs}$\xspace}
\newcommand{\etheo}{$\epsilon_{\rm ff,100}$\xspace}
\newcommand{\eff}{$\epsilon_{\rm ff}$\xspace}

\newcommand{\avir}{$\alpha_{\rm vir}$\xspace}

\usepackage[normalem]{ulem}
\title[Star formation efficiency in the MW]{Cosmic evolution of the star formation efficiency in Milky Way-like galaxies}

\author[Segovia Otero et al.]
{Álvaro Segovia Otero$^{1}$\thanks{alvaro.segovia@astro.lu.se}, %
Oscar Agertz$^{1}$,
Florent Renaud$^{1,2,3}$,
Katarina Kraljic$^{2}$,
Alessandro B. Romeo$^{4}$,
\newauthor and Vadim A. Semenov$^{5}$
\\
$^{1}$Lund Observatory, Division of Astrophysics, Department of Physics, Lund University, Box 43, SE-221 00 Lund, Sweden\\
$^{2}$Observatoire Astronomique de Strasbourg, Universit\'e de Strasbourg, CNRS UMR 7550, F-67000 Strasbourg, France\\
$^{3}$University of Strasbourg Institute for Advanced Study, 5 all\'ee du G\'en\'eral Rouvillois, F-67083 Strasbourg, France\\
$^{4}$Department of Space, Earth and Environment, Chalmers University of Technology, SE-41296 Gothenburg, Sweden\\
$^{5}$Center for Astrophysics $|$ Harvard \& Smithsonian, 60 Garden St, Cambridge, MA 02138, USA
}

\date{Accepted XXX. Received YYY; in original form ZZZ}

\pubyear{2024}

\begin{document}
\label{firstpage}
\pagerange{\pageref{firstpage}--\pageref{lastpage}}
\maketitle

\begin{abstract}
Current star formation models are based on the local structure of the interstellar medium (ISM), yet the details on how the small-scale physics propagates up to global galactic-scale properties are still under debate. To investigate this we use \vintergatan, a high-resolution (20 pc) cosmological zoom-in simulation of a Milky Way-like galaxy. We study how the velocity dispersion and density structure of the ISM on 50-100 parsec scales evolve with redshift, and quantify their impact on the star formation efficiency per free-fall timescale, \eff. During epochs of starburst activity the star forming ISM can reach velocity dispersions as high as $\sim 50 \kms$ for the densest and coldest gas, most noticeable during the last major merger event happening at $1.3 < z < 1.5$. After a phase dominated by mergers ($1 < z < 5$), \vintergatan transitions into a secularly evolving state where the cold neutral ISM typically features velocity dispersion levels of $\sim 10 \kms$. Despite strongly evolving density and turbulence distributions over cosmic time, the local \eff at the resolution limit is found to change by only a factor of a few: from median efficiencies of 0.8\% at $z>1$ to 0.3\% at $z<1$. The corresponding mass-weighted average shows a universal $\langle \epsilon_{\rm ff} \rangle \approx 1\%$, originating from an almost invariant distribution of virial parameters in star forming clouds, where changes in gas densities and turbulence levels are coupled such that the kinetic-to-gravitational energy ratio remains close to constant. Finally, we show that a \textit{theoretically} motivated instantaneous efficiency such as \eff is intrinsically different to its \textit{observational} estimates adopting tracers of star formation e.g. H$\alpha$. As signatures of the physical conditions that trigger star formation can be lost on short ($\sim$ 10 Myr) timescales, we argue that caution must be taken when constraining star formation models from observational estimates of \eff.

\end{abstract}


\section{Introduction}

Current star formation theories strive to connect, and track across cosmic time, the local properties of star-forming regions ($\lesssim 100 \pc$), the global parameters characterising galaxies ($>$ tens of kpc), and their cosmological environment. Nearby star-forming spiral galaxies follow the canonical main sequence (MS) and Kennicutt-Schmidt (KS) relations: at a given stellar mass, disc galaxies with larger molecular gas surface densities present greater star formation rate (SFR) surface densities, depleting their gas reservoirs on timescales of $\sim$ few Gyr \citep[gas depletion times are estimated as $\tau_{\rm dep} = M_{\rm gas}/{\rm SFR}$,][]{kennicutt98,bolatto08,leroy13,speagle14}. The absence of major mergers combined with low gas fractions ($f_{\rm g} \lesssim 10\%$ for $M_\star\gtrsim 10^{10}~{\rm M_{\odot}}$, e.g. \citealt{Scholte2024}) allow these isolated systems to develop stable rotationally-supported thin discs \citep{brinchmann04,wuyts11,wisnioski15}, where star formation on galactic scales is regulated by gas accretion, disc dynamics and instabilities, and feedback processes.

In the Local Universe, it is mainly within giant molecular clouds (GMCs) that the ISM is shielded from background radiation, reaching the required density and temperature conditions to trigger star formation \citep[for a review, see][]{McKeeOstriker2007}. CO observations of GMCs in the Milky Way have revealed cloud scaling relations that relate their masses, sizes and velocity dispersions \citep{larson81,Solomon1987,heyer09,mivilledeschenes17}. These have typical molecular gas surface densities of $\Sigma_{\rm gas} \approx 100 \Msunpcs$, effective radii of $R \sim 10-100 \pc$, and velocity dispersions of $\sigma_{\rm gas} \sim 1-10 \kms$. These properties translate into cloud virial parameters of $\alpha_{\rm vir} \approx 1-10$ \citep{leroy16,mivilledeschenes17,sun22}, where \avir is defined as the ratio between the kinetic to gravitational energy such that $\alpha_{\rm vir}=2 E_{\rm kin}/|E_{\rm grav}|$. A constant \avir of $\gtrsim$1 is often understood as GMCs reaching dynamical equilibrium, although external cloud confinement by thermal and turbulent pressure complicates this picture \citep[e.g.][]{Elmegreen1989,grisdale18}. Star formation under these conditions is inefficient, which is commonly quantified using the star formation efficiency per free-fall time parameter \eff, i.e. the fraction of stars formed from the total gas mass of a GMC on a free-fall timescale $t_{\rm ff}=\sqrt{3 \pi / 32 G \rho }$. Observations of star-forming GMCs show \eff distributions with a universal average of 1\%, followed by a spread of several orders of magnitude \citep[see review by][]{Krumholz2019}. Preliminary systematic changes in \eff and \avir have been found as a function of galactic environment (bulge, bars, spiral arms, disc), but the scatter around these parameters is dominated by intrinsic cloud-to-cloud differences \citep{sun20a,sun20b,rosolowsky21}. The way in which gas densities and velocity dispersions in the ISM are affected by environmental factors, and whether this has a significant impact on the local star formation properties, still remains unclear even at low redshift.

At high redshifts ($z>1$), star-forming galaxies are predominantly rotating discs, and their high gas fractions make them gravitationally unstable to clump formation \citep[for reviews see][]{forstershreiberwyuts20,saintonge22}. Such a clumpy ISM is subject to strong feedback from enhanced SFRs, along with more frequent galactic interactions \citep{duncan19}. All of this sets elevated velocity dispersions at high redshifts, as measured from kinematics of molecular, atomic, and ionised gas tracers \citep[][and references therein]{ubler19}. Compared to spiral disc galaxies at present day, average gas depletion times in high-$z$ galaxies are found to be shorter \citep[\tdep$\geq 500$ Myr, see][]{tacconi18,tacconi20}. In contrast to low redshift discs, the global \tdep in starburst galaxies is around an order of magnitude shorter \citep[30-300 Myr,][]{daddi10,rodighiero11}, making them a natural testbed for star formation physics in more extreme ISM conditions. Starbursts can be driven by mergers, but can also be triggered by secular processes in isolated gas-rich turbulent discs \citep{ciesla23}. 

The star-forming gas clumps in starbursts observed to date are denser and more turbulent than GMCs in Local spirals, nevertheless, they also show $\alpha_{\rm vir} \approx 1-10$, albeit with a large scatter \citep[][see also \citealt{krahm24} for GMCs in the overlap region between the Antennae galaxies]{leroy15,rosolowsky21,mirka23}. Unfortunately, observations of gas clumps in such objects are scarce and limited to bright mergers in the Local Universe or lensed high-redshift galaxies \citep{mirka19}. It is therefore uncertain whether ISM conditions at high redshift or in extreme environments such as starbursts imply efficient local star formation, i.e. a higher \eff.

Mapping the density and turbulent structure of the ISM to its star formation properties has been the goal of theoretical and computationally-driven studies in the last couple of decades. Pioneering analytical work base the star formation process on the theory of supersonic isothermal gas \citep{km05,pn11,hc11,fk12,burkhart18}. These models, further calibrated on controlled magnetohydrodynamic (MHD) simulations of GMCs \citep[5-500 pc in size,][]{grudic18,grudic19} and highly resolved ISM boxes \citep[$\sim$ few pc in length,][]{federrath10,federrath13,federrath15}, can predict the commonly observed average $\langle \epsilon_{\rm ff}\rangle \approx 1 \%$ even though they locally present large \eff variations \citep{krumholz12}. More realistic set-ups face the challenge of resolving the large dynamic range that connects the inner structure of GMCs and their galactic and extra-galactic environment in isolated disc or cosmological simulations respectively. To that aim, sub-grid star formation recipes are either based on the aforementioned analytical prescriptions where \eff directly depends on the local gas density and velocity dispersion \citep{semenov16,trebitsch17,kretschmer20,arturo21}, or have a fixed \eff typically in the range of 1-100 percent \citep{renaud13,hopkins18,grisdale18,grisdale19}. These simulations reveal a complex interplay among the extra-galactic and galactic environment, star formation models, and stellar feedback prescriptions \citep[e.g.][]{agertz11,agertzkravtsov15,semenov18}. For this reason, different star formation models can yield significant differences in ISM and GMC properties \citep[e.g.][]{grisdale17,grisdale18}, but be degenerate in terms of integrated quantities on galactic scales such as SFRs or global gas depletion times \citep[][]{hopkins13c}. 

In this paper we use \vintergatan \citep{vinter1,vinter2,vinter3}, a cosmological zoom-in simulation of a Milky Way-like galaxy to investigate the evolution of the ISM and local star formation properties, i.e. \eff as a function of redshift. In our simulation, the \citet{pn12} effective model connects the density and velocity disperison of the ISM with \eff, which we further compare to Local Universe GMC observations from the PHANGS collaboration \citep[][]{schinnererleroy24}. This work is a follow up of \citet{aso22} and \citet{renaud22}, where we demonstrated the crucial role played by galaxy mergers and morphological transformations in setting the global gas depletion time. Here we focus on investigating the physics on smaller ($\lesssim 50-100$ parsec) scales and how it connects to the global properties of a galaxy across cosmic time. Section \ref{sec:method} briefly summarizes the simulation set-up, Section \ref{sec:results} highlights the connection between the cosmological environment and the local star formation properties of Milky-Way like galaxies, and Section \ref{sec:discussion} and \ref{sec:conclusions} conclude with the main takeaways of this work and contextualise them with respect to other available models for star formation.

\begin{figure*}
    \centering
    \includegraphics[width=0.95\textwidth]{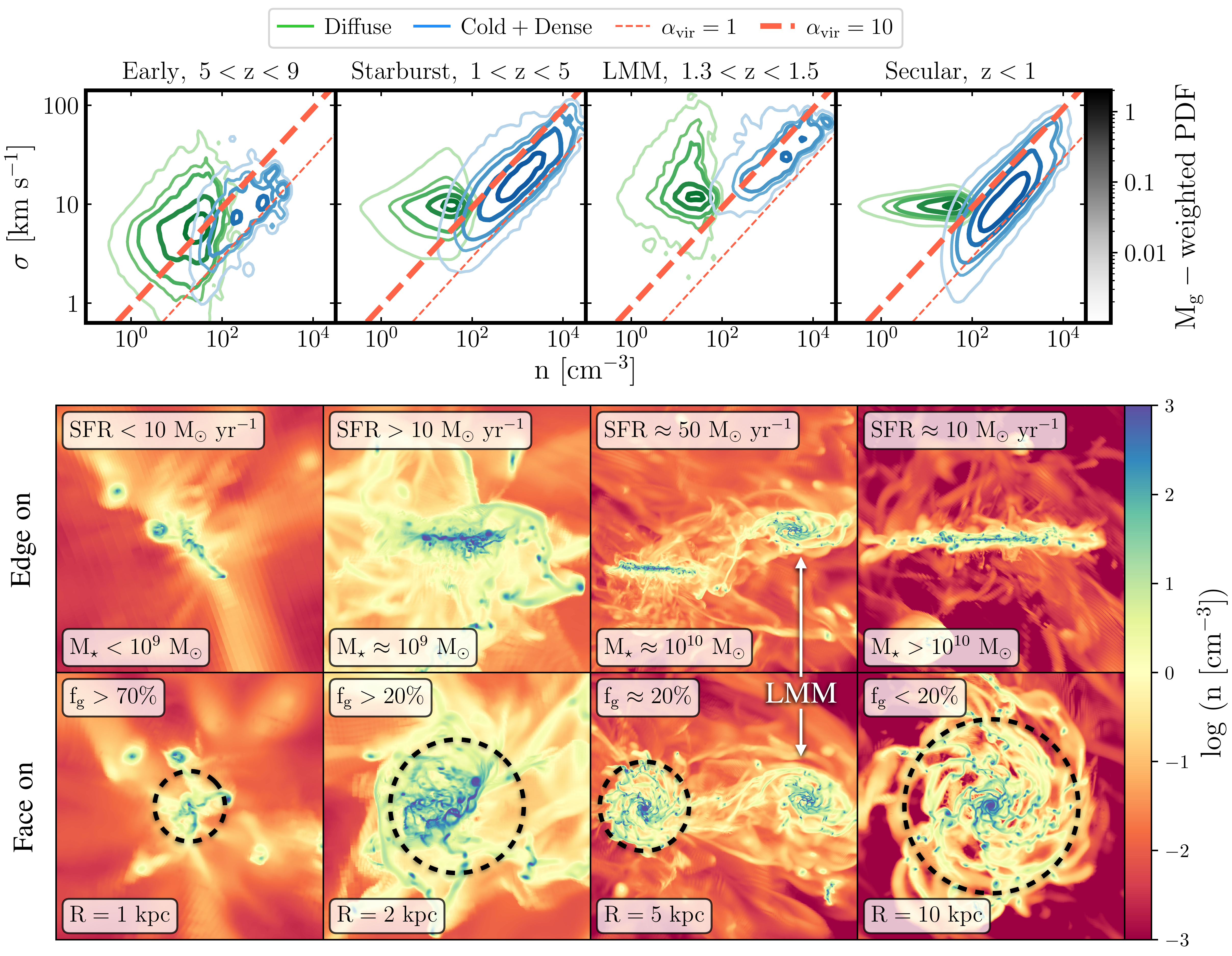}
    \caption{\textit{\textbf{Top:}} Response of the ISM in \vintergatan as a function of redshift to its extra-galactic environment. For every redshift interval, each panel shows the distribution of the density $n$ and velocity dispersion $\sigma$ calculated from gas in (50 pc)$^3$ cubes. Contours represent the mass-weighted 2D histogram with probability densities of 0.6, 0.3, 0.1, 0.05, and 0.01 from thickest and darkest to thinnest and brightest contours respectively. Two sets of contours are presented: \textit{cold+dense} (blue) for gas $T<100$ K and $n>100 \cc$ (star-forming gas according to Equation \ref{eq:sf_pn12}); \textit{diffuse} (green) for gas $T < 10^{4}$ K (excluding \textit{cold+dense} gas). Dashed red lines of constant \avir are included (Equation \ref{eq:avir}), computed on 50 pc scales for $\alpha_{\rm vir} = 1$ (thin) and $\alpha_{\rm vir} = 10$ (thick). \textit{\textbf{Bottom:}} Snapshots of the evolution of \vintergatan. Each column represents the edge- and face-on views of the galaxy, colour-coded by its gas density at every cosmic epoch: \textit{Early} ($5 < z < 9$), \textit{Starburst} ($ 1 < z < 5$) including the \textit{LMM} event from the first pericenter passage ($z \approx 1.5$) to the final coalescence of the merger ($z \approx 1.3$), and \textit{Secular} ($z < 1$). Included in each panel is the SFR, computed from the mass of stars younger than 100 Myr, the total stellar mass $M_{\star}$, the gas fraction $f_{g}$, and twice the half radius calculated accounting for the spatial distribution of stars younger than 100 Myr (and outlined in black dashed circles on the bottom row panels). \textit{\textbf{Takeaway:}} Density and velocity dispersion go hand in hand, specially for cold and dense gas, and respond to the cosmological environment: mergers stir the ISM towards high $\sigma$ and $n$, when a more isolated disc has more lower turbulent levels and density values.}
    \label{fig:threephaseism}
\end{figure*}

\section{Methodology}
\label{sec:method}
The \vintergatan cosmological zoom-in simulation was run using the hydrodynamics+{$N$}-body code \ramses \citep{teyssier02}, assuming $\Lambda$-cold dark matter ($\Lambda$CDM) cosmological parameters\footnote{$H_0 = 70.2 \kms$ Mpc{$^{-1}$}, $\Omega_{\rm m} = 0.272$, $\Omega_{\Lambda} = 0.728$, $\Omega_{\rm b} = 0.045$}. The initial conditions correspond to the 'm12i' halo of the \agora project (\citealt{kim14}, \citealt{kim16}, \citealt{santi21}, see also \citealt{Wetzel2023}), generated with the \music code \citep{hannabel11}. Within a periodic box of 85 Mpc and $512^3$ dark matter particles, the progenitor Lagrangian region is selected to be $3R_{200,{\rm m}}$ around the halo at $z = 0$, where the virial radius and mass with respect to the mean cosmic background density at $z=0$ are $R_{200,{\rm m}} = 334~{\rm kpc}$ and $M_{200,{\rm m}} = 1.3 \times 10^{12}\Msun$ respectively. Its resolution was further enhanced to reach a dark-matter particle resolution of $3.5 \times 10^4 \Msun$, and a mass and spatial resolution of $7070 \Msun$ and $\lesssim 20 \pc$ respectively for the gas. A more exhaustive explanation of the technical aspects of the simulation can be found in \citet{vinter1}. Here, we briefly emphasise on the relevant models applied to describe the galaxy formation physics within \vintergatan.

Star formation is treated as a Poisson process on a cell-by-cell basis \citep[see][]{agertz13}, where the star formation rate density {$\Dot{\rho}_{\star}$} follows

\begin{equation}
    \label{eq:sf_schmidt}
    \Dot{\rho}_{\star} = \epsilon_{\rm ff} \dfrac{\rho_{\rm g}}{t_{\rm ff}}, \quad {\rm with} \quad \rho_{\rm g}>100 \cc, \quad {\rm and} \quad T_{\rm g} < 100 \ {\rm K}.
\end{equation}
Here, $\tff$ is the free-fall timescale and \eff is the star formation efficiency per free-fall time from \citet{pn12} (henceforth, PN12). Star particles with an initial mass of {$10^4 \Msun$} form when gas is colder and denser than the density and temperature thresholds. These conditions are set to prevent thermally-supported gas from forming stars, a choice that has a minor impact on the overall star formation history as the free-fall timescales associated to such diffuse gas are long and do not contribute to the global SFR.

In PN12 they simulate star formation in supersonically turbulent, magnetized gas with a wide range of initial conditions and find that the star formation efficiency per free-fall time is well approximated by

\begin{equation}
    \label{eq:sf_pn12}
    \epsilon_{\rm ff} = \epsilon_{\rm w} \ \exp \left(-1.38 \sqrt{\alpha_{\rm vir}}\right) = \epsilon_{\rm w} \ \exp \left(-1.6 \dfrac{\tff}{\tturb}\right).
\end{equation}
The normalisation constant $\epsilon_{\rm w}$, here set to $0.5$, accounts for processes such as mass loss via proto-stellar jets. The virial parameter is defined as

\begin{equation}
    \label{eq:avir}
    \alpha_{\rm vir} = \dfrac{2 E_{\rm kin}}{|E_{\rm grav}|} = \dfrac{15 \sigma_{1D}^2}{\pi G \rho_{\rm g} L^2} = 1.35 \left(\dfrac{t_{\rm ff}}{t_{\rm turb}}\right)^2,
\end{equation}
where
\begin{equation}
    \label{eq:tcr}
    t_{\rm turb} = \dfrac{L}{2 \sqrt{3}\sigma_{\rm 1D}}
\end{equation}
is commonly referred to as the turbulence crossing time, which is on the order of the timescale for turbulent energy dissipation in the case of driven supersonic turbulence \citep[][]{maclow98,Stone1998,maclow99}. $L$ is chosen to be the cell size ($\lesssim 20$ pc), and $\sigma_{\rm 1D}$ is the one-dimensional velocity dispersion calculated as the standard deviation of each component of the velocity vector independently for all 6 adjacent cells, i.e. effectively computed on $\sim 50$ pc scales, $\sigma_{\rm 1D}^2 = (\sigma_{\rm x}^2 + \sigma_{\rm y}^2 + \sigma_{\rm z}^2)/3$.

The star formation model in PN12 is exclusively virial parameter-dependent. Alternatives exist, e.g. with additional Mach number dependencies \citep[for a comprehensive discussion see][]{fk12}, but have not been directly implemented in this work. In Section \ref{sec:discussion} we comment on the potential implications of using different prescriptions for \eff. It is also worth mentioning that even though sophisticated, these star formation laws do not directly account for the galactic scale environment surrounding the star-forming region, hence not accounting for mechanisms such as shear and large-scale compression.

Each star particle represents a single stellar population with a universal initial mass function \citep[IMF,][]{chabrier03} and age-, mass-, and gas/stellar metallicity-dependent feedback processes such as radiation pressure, stellar winds, core collapse, and type Ia supernovae calibrated on the \starburst code \citep{raiteri96,leitherer99,kimostriker15}. Gas metallicity is initialised from a floor of $Z = 10^{-3}$ Z{$_{\odot}$} in the zoomed-in region\footnote{This metallicity floor accounts for unresolved population III stars \citep{wise12,agertz20}.} and subsequently enriched with iron (Fe) and oxygen (O) by supernovae, where we adopt the yields from \citep{woosleyheger07}. This allows for metallicity-dependent cooling for $T<10^4$ K \citep{rosenbergmas95} and $10^{4} < T < 10^{8.5}$ K gas \citep{suthdopita93}. Gas heating from a UV radiation background was added assuming reionisation starting at $z = 8.5$ \citep{haardmadau96,courtyalimi04,auberteyssier10}.

To extract reliable information from our data, we use robust statistics \citep{muller00,romeo23}. Namely, the median and median absolute deviation (MAD) provide reliable information on the 'central value' and 'width' of the sample even when a large fraction of them are outliers. The robust 1$\sigma$ scatter is derived as MAD/0.6745. We further emphasize that using the median rather than the mean is convenient under logarithmic transformations given that ${\rm med} \left[\log \left(X\right)\right] = \log \left[{\rm med} \left(X\right)\right]$. The robust scatter in logarithmic scale is approximated to $\sigma_{\log \left(X\right)} \approx \sigma_{X} / \left[X\ln(10)\right]$, which is more representative in the limit where $\sigma_{X}/X \ll 1$.

\section{Results}
\label{sec:results}

\vintergatan proposes a formation channel general to nearby star-forming spirals that reproduces Milky Way features at {$z \approx 0$} \citep{vinter1,vinter2,vinter3}. \citet{aso22} demonstrate that Milky Way-like galaxies experience epochs of starburst or more quiescent star formation activity depending on the global gas depletion time of the cold ISM. They identify three evolutionary stages with order of magnitude changes in \tdep: an \textit{early} stage at high redshift ($5 < z < 9$) with $\tau_{\rm dep} \sim 1 \Gyr$; a \textit{starburst} phase ($1 < z < 5$) when \tdep drops by an order of magnitude driven by mergers impacting on an assembled galactic disc; a \textit{secular} stage ($z < 1$) when \tdep increases to a few Gyr in the absence of mergers. In the following, we explore how the density and turbulent structure of the ISM evolve with time depending on these epochs, and asses how much this affects the local star formation properties in the galaxy. Note that the word \textit{local} is used in this work to refer to ISM properties derived on GMC scales when speaking about observations and for parameters computed on scales comparable to the size of gas cells in simulations. Conversely, the word \textit{global} refers to galactic scales.

\subsection{ISM density and velocity dispersion across cosmic time}

Figure \ref{fig:threephaseism} shows the ISM in the aforementioned evolutionary stages of \vintergatan. An additional panel representing the last major merger (LMM) exemplifies the properties of the ISM from its first pericenter passage ($z \approx 1.5$) to its final coalescence with the main galaxy ($z \approx 1.3$). Maps showing the gas density of the main galaxy (face- and edge-on), its environment, and its main global properties at each phase are on the bottom. In the top panels, contours show the density and velocity dispersion levels of the ISM, computed by stacking simulation snapshots with an output resolution of $100 \Myr$ within each redshift interval. The dashed red lines mark regions of constant virial parameters of $\alpha_{\rm vir} =1$ (thin) and $\alpha_{\rm vir} = 10$ (thick) using Equation \ref{eq:avir}.

For every output, we select a spherical volume with radius equal to three times the stellar half-mass radius of the galaxy\footnote{Galactic centers in each output are found by running a shrinking sphere algorithm, and half-mass radii are calculated from the mass of stars younger than 100 Myr within the virial radius of \vintergatan at each epoch.} and split it in cubes of 50 pc in size. Within each cube we divide the gas into two phases by applying density and temperature cuts:

\begin{itemize}
    \item The \textit{cold+dense} ISM (blue contours) corresponds to gas that fulfills the star formation criteria from Equation \ref{eq:sf_schmidt}, i.e. above $n > 100 \cc$ and below $T < 100$ K. This phase traces molecular species \citep[]{KlessenGlover2016}. Densities are computed by dividing the corresponding mass of \textit{cold+dense} gas by the volume of the 50 pc cube, and $\sigma$ represents the one-dimensional mass-weighted standard deviation of the velocities in 50 pc cubes with more than one \textit{cold+dense} gas cell.
    
    \item The \textit{diffuse} ISM (green contours) includes all gas cells at temperatures of $T < 10^4 \K$, excluding those with \textit{cold+dense} gas properties. These density and temperature ranges are  typical of the warm neutral or ionised medium traced by neutral hydrogen (HI) and recombination lines such as H$\alpha$. Its densities and velocity dispersions have been computed in the same way as for the \textit{cold+dense} gas, minding the new temperature and density cuts.
\end{itemize}

\begin{figure*}
    \centering
    \includegraphics[width=0.835\textwidth]{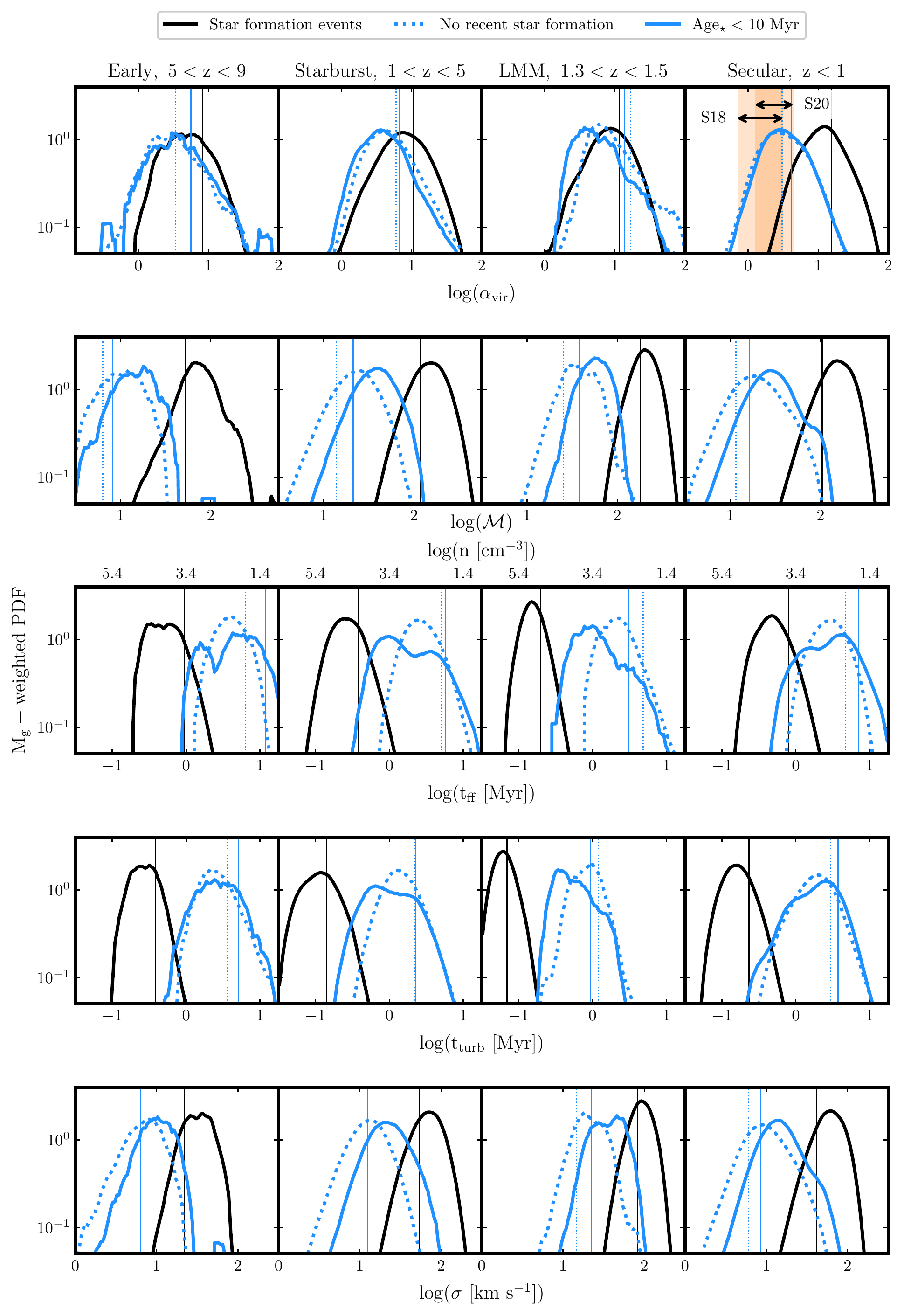}
    \caption{Gas mass-weighted PDFs of \avir, $\mach$, $\tff$, the turbulent timescale $\tturb$, and velocity dispersion at each cosmic epoch in \vintergatan. PDFs plotted in solid black correspond to \textit{SF-events} while those in blue represent \textit{cold+dense} gas. Curves in solid blue show ISM properties evaluated on 50 pc cubes with star particles younger than 10 Myr. Those in dotted blue PDFs characterise the ISM with no recent star formation. Additionally, vertical lines correspond to the median values of each distribution with same colour and line style. The orange rectangles in the top right panel represent the scatter (robust standard deviation, see last paragraph in Section \ref{sec:method}) in \avir values evaluated on 80 pc in \citet{sun18} and 90 pc in \citet{sun20b}. Black arrows are just included to clarify the extent of the scatter. \textit{\textbf{Takeaway:}} Even if the density and turbulence of the ISM on 50 pc scales changes with cosmic epoch, the virial parameter shows a roughly constant distribution as a function of redshift.}
    \label{fig:1dh}
\end{figure*}

The velocity dispersion and density distributions for both \textit{diffuse} and \textit{cold+dense} gas phases evolve with cosmic epoch. This evolution happens at a constant virial parameter (see Section \ref{sec:avir}), with $\alpha_{\rm vir} \geq 10$ for \textit{diffuse} gas and at $1 < \alpha_{\rm vir} < 10$ for \textit{cold+dense} gas. In fact, \textit{cold+dense} gas shows a positive correlation between the velocity dispersion and density with $\sigma \propto n^{0.5}$, in agreement with previous studies of disc galaxies \citep{semenov16}. Such a scaling is compatible with the Larson scaling relations \citep{larson81,grisdale18} when a \emph{fixed spatial scale} is considered. \textit{Diffuse} gas has less inertia and is easier to accelerate, making it more susceptible to stellar feedback \citep[][]{timmy22} and the extra-galactic environment. Thus, it deviates from the $\sigma-n$ relation, with substantially more scatter, specially at $z > 1$.

During the \textit{early} stage ($5<z<9$), most of the gas is found at velocity dispersion values of $\sigma \leq 30 \kms$, which is in line with CII observations of galaxies in the $10^8 \leq M_{\star} \leq 10^{10} \Msun$ mass range at these redshifts \citep[left panel in Figure \ref{fig:threephaseism}, see also][]{pope23,posses23}. Although gas-rich and immersed in a merger-dominated environment, SFRs in \vintergatan are the lowest in its star formation history \citep[\citealt{aso22}, for observations of gas fractions see][and \citealt{fudamoto22,leethochawalit23,trussler23} for SFRs of coeval systems]{heintz22,aravena23}. It is not until the galaxy enters the \textit{starburst} stage ($1<z<5$) that the highest velocity dispersion levels are reached, especially in the case of the \textit{LMM} event for both gas phases (central panels in Figure \ref{fig:threephaseism}). This is in agreement with observed starbursting discs slightly more massive than \vintergatan at similar redshifts \citep[$\sigma \approx 20-80 \kms$ also from CII measurements,][]{herreracamus22,romanolveira23,parlanti23,rizzo23}. With the assembly of a galactic disc at {$z \approx 4.8$}, mergers tidally compress gas, triggering star formation on short depletion times due to an excess of gas in the \textit{cold+dense} phase \citep{aso22,renaud22}. Nevertheless, even if an excess of gas at high densities leads to high SFRs (which inevitably goes hand-in-hand with increased stellar feedback), disentangling the roles of feedback- and gravity/compression-driven turbulence is challenging in cosmological zoom-in simulations \citep[but see][who tell both effects apart using high time-resolution simulations of idealised mergers]{renaud14}. In the absence of major mergers (\textit{secular} stage, $z \leq 1$), and with the depletion of gas reservoirs, SFRs and velocity dispersion values decrease most notably in the \textit{diffuse} gas \citep[$\sigma < 10 \kms$ for HI observations in Local spirals,][]{tamburro09,eibensteiner23}. Consequently, its scatter also decreases substantially.

\begin{figure*}
    \centering
    \includegraphics[width=\textwidth]{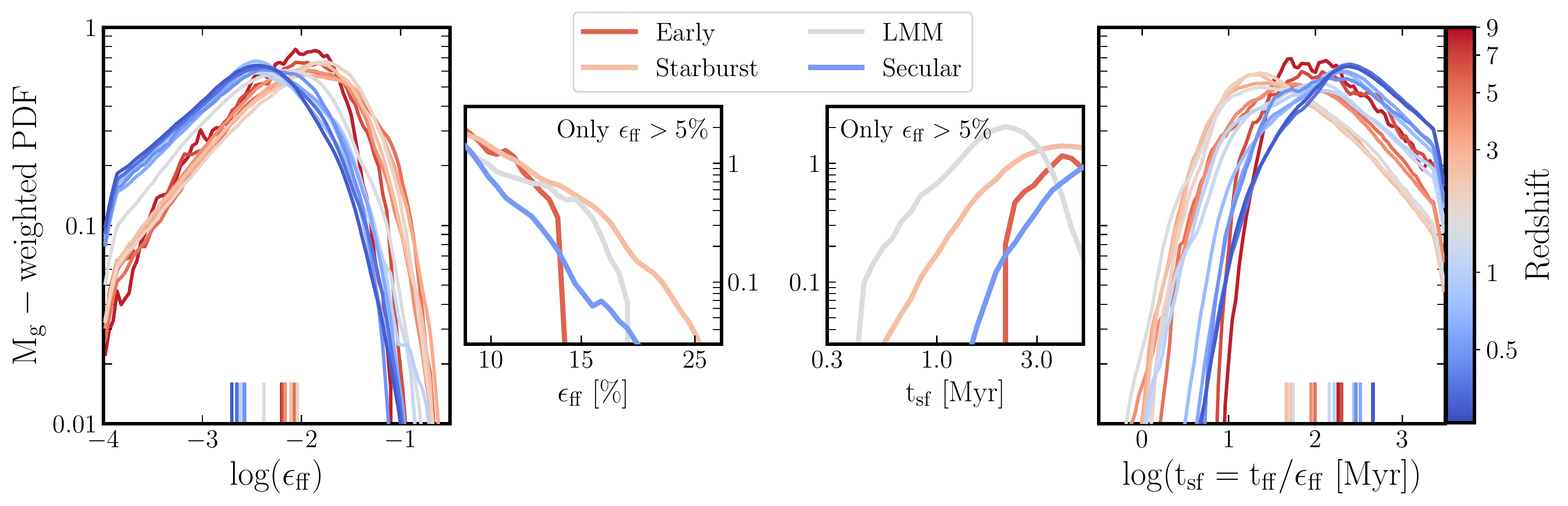}
    \caption{Cosmic evolution of the \eff gas mass-weighted PDFs (\textit{left}) and local star formation timescales (\textit{right}) in 15 redshift bins. The on-the-fly \eff is calculated with Equation \ref{eq:sf_pn12} \citep{pn12}, using \textit{SF-events} properties as input. Small line markers at the bottom the left and rightmost panels indicate the evolution of the median \eff and $t_{\rm sf}$ respectively. Two additional panels display the re-calculated PDFs for only cells with $\epsilon_{\rm ff}>5\%$ (\textit{center left}). For clarity, 4 colour-coded PDFs show the averaged distributions for each cosmic epoch. \textit{\textbf{Takeaway:}} $\langle \epsilon_{\rm ff} \rangle \approx 1\%$ is constant throughout cosmic time, and therefore the local timescales for star formation are mostly dependent on the changes in the gas density PDFs, through $\tff$.}
    \label{fig:effevolution}
\end{figure*}

\subsection{Properties of the cold and dense ISM}
\label{sec:avir}
Densities reached in \vintergatan imply free-fall timescales at least a factor of ten shorter than the output frequency of the simulation ($\sim 100 \Myr$). Thus, to get insight into the actual ISM properties upon star formation we use the \textit{star formation events} (hereafter \textit{SF-events}) gas. This data-set is different to the previous two gas phases in that they have not been extracted directly from the low-cadence outputs of the simulation. \textit{SF-events} properties are logged on the fly from gas cells that have undergone star formation: densities are directly read from the star-forming cell ($\lesssim 20 \pc$), and velocity dispersion values are calculated using the velocity vectors of the six adjacent cells ($\sim 50 \pc$). \textit{SF-events} gas is represented in solid black in Figure \ref{fig:1dh}, revealing very turbulent ($\geq 50 \kms$), dense ($\geq 10^{4} \cc$), and cold ($\approx 10 \K$) properties that resemble those that make molecules such as HCN, HCO$^+$, NH$_3$, CH, CN, and CS \citep[]{KlessenGlover2016,gallagher18b,wilson23}.

From this point on we focus our analysis on the \textit{cold+dense} and \textit{SF-events} gas in order to connect the ISM structure with its star formation properties at each cosmic epoch. Figure \ref{fig:1dh} shows the evolution of \avir (Equation \ref{eq:avir}), Mach number\footnote{The expression for the Mach number is $\mathcal{M} = \sqrt{3} \sigma_{\rm 1D} / c_{\rm s}$, where $c_{\rm s}$ is the thermal sound speed determined by the temperature of gas in the cell $T$, the Boltzmann constant $k_{\rm B}$, a mean molecular weight of $\mu=1.3$, and the mass of hydrogen $m_{\rm H}$ so that $c_{\rm s} = \sqrt{k_{\rm B}T/\mu m_{\rm H}}$. Note that the sound speed for both \textit{cold+dense} and \textit{SF-events} gas is $c_{\rm s} \lesssim 1 \kms$, making turbulence supersonic for all measured values of the velocity dispersion.}, $\tff$, $\tturb$ (Equation \ref{eq:tcr}), density, and velocity dispersion for \textit{SF-events} gas in black and \textit{cold+dense} gas in blue. \textit{Cold+dense} gas is further split in two: gas within 50 pc cubes containing stars younger than 10 Myr in solid blue; gas cubes without star particles younger than 10 Myr in dotted blue. It is worth noting that the velocity dispersion values in both \textit{cold+dense} and \textit{SF-events} gas are computed at approximately the same scales, but other properties such as $\tff$ (or $c_{\rm s}$) are derived from the density (or mass-weighted average temperature) on 50 pc cubes compared to reading off the star-forming cell properties on $\lesssim 20$ pc before spawning a star particle. We emphasize that our goal here is to: $i)$ analyse the properties of the ISM at the resolution limit of the simulation where the PN12 star formation model actually comes into play; $ii)$ compare the distributions of \textit{SF-events} and \textit{cold+dense} properties; $iii)$ compare \vintergatan to GMC scaling relations from the PHANGS collaboration, who provide GMC properties on a variety of scales \citep[45-150 pc,][]{rosolowsky21,sun23,schinnererleroy24}.

The most striking result in Figure \ref{fig:1dh} is the difference between the \textit{SF-events} and the \textit{cold+dense} gas distributions. Note that the distributions are normalised, so \textit{SF-events} gas actually conforms a smaller sample of cells representing the more extreme regions of the $\sigma-n$ space. An order of magnitude shift to shorter $\tff$ and $\tturb$ dynamical timescales is seen, highlighting the importance of capturing these short-lived events, and implying that \textit{SF-events} gas is indeed a biased subset of denser and more turbulent gas intimately connected to stellar feedback. To explore this idea, we compare \textit{cold+dense} gas with and without recent star formation (solid and dotted blue in Figure \ref{fig:1dh}). We find that gas embedding stars younger than $10 \Myr$ reach denser and more turbulent states, i.e. stronger resemblance to \textit{SF-events} distributions. The same trends are recovered when defining recent star formation as stars younger than 5 Myr and 100 Myr (timescales probed by H${\alpha}$ and UV observations respectively), with stronger shifts between recent and past star forming ISM the older the stellar age cut is.

Looking at the \textit{cold+dense} distributions, the shortest $\tff$ and $\tturb$ are achieved during the \textit{LMM} event compared to the other epochs (third and fourth rows in Figure \ref{fig:1dh}), expected from the evolution of the corresponding density and velocity dispersion distributions. Nevertheless, both dynamical timescales conspire to render an \avir PDF that is roughly constant in all four cosmic epochs (top row in Figure \ref{fig:1dh}), with median values of $\alpha_{\rm vir} \approx 4.4$ and robust scatter of 3.3. Approximately stationary \avir distributions with time is also noticed for \textit{SF-events} gas, which overlap substantially with \textit{cold+dense} gas ones despite the stark differences between them. This is particularly the case at $z>1$, with both \textit{SF-events} and \textit{cold+dense} \avir distributions showing marginally larger median values. In merger-driven environments at these redshifts, higher \avir values are observed in recent simulations by \citet{he23}, yet CO observations of GMCs in mergers and lensed turbulent discs at $z>1$ cannot conclusively estimate a systematic change. Some studies point towards $\alpha_{\rm vir} \approx 1$ in both mergers and isolated galaxies \citep{mirka19,mirka23}, whereas others report increased \avir values in starbursts, low density regions, galactic center environments, and even mergers \citep{leroy15}. For the particular case of the Antennae galaxies, the order of magnitude increase in physical resolution to $\sim 10$ pc from \citet{krahm24} compared to \citet{wei12} renders an average of $\langle \alpha_{\rm vir}\rangle = 13.80 \pm 7.84$.

In the absence of mergers during the \textit{secular} phase, the median \avir is around a factor of three larger in \textit{SF-events} gas with respect to \textit{cold+dense}. Such moderate shift to lower virial parameters in \textit{cold+dense} gas is due to lower velocity dispersion values increasing $\tturb$, and is in good agreement with CO-based observations of GMCs in nearby galaxies\footnote{Our equation for \avir is almost identical to that of \citet{sun18} and \citet{sun20b}, except that in their work they use the projected gas surface density, $\Sigma_{g}=\pi R^{2}$, where $R$ is the depth of the line of sight. They also add a geometrical factor $f$ that corrects for an assumed radial density profile of CO inside their spherical GMCs so that $f=(1-\gamma/3)(1-2\gamma/5)$, where $\gamma$ is the slope of the power-law of the profile \citep{bertoldimckee92,rosolowskyleroy06}.} \cite[highlighted in orange boxes on the top right panel in Figure \ref{fig:1dh},][]{sun18,sun20b,rosolowsky21}. On the other hand, \textit{SF-events} experiences an increase in \avir due to the fact that less gas is compressed to dense states, creating longer $\tff$ on average. As we mentioned previously, if we take \textit{SF-events} to resemble dense gas tracers, its \avir values overestimate the HCN-based virial parameters \citep{gallagher18a,empire19,neumann23}, which agree better with CO-based observations.

Additionally, we have checked that our results are robust to small changes to the spatial scale by repeating the analysis using 100 pc cubes, with no change to the main conclusions drawn here.

\begin{figure*}
    \centering
    \includegraphics[width=0.9\textwidth]{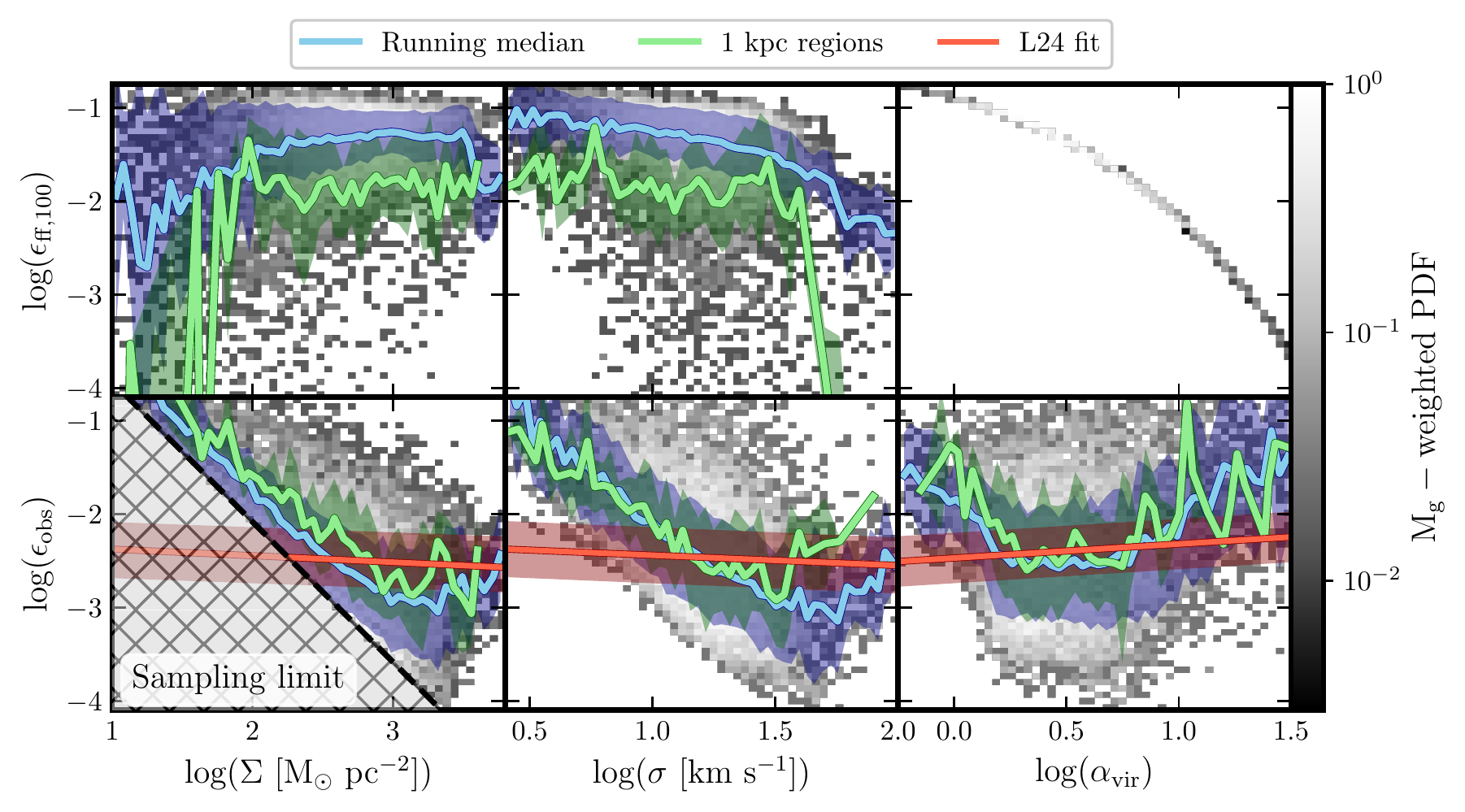}
    \caption{Theoretical (\textit{top row}) and observational (\textit{bottom row}) estimates of the star formation efficiencies per free-fall time as a function of $\Sigma$, $\sigma$ and \avir. Only \textit{cold+dense} gas properties in outputs at $z<1$ are used. The running median of the 2D histograms analysed on a grid of cubes of 100 pc in size is shown in blue, with its robust standard deviation as blue shaded band (see Section 3.2). In solid green we have the PHANGS-like running median also with its associated robust standard deviation band evaluated over on a grid of 1 kpc cubes (see Section 3.4 for a more thorough explanation on how to calculate PHANGS-like properties). In addition, the bottom row contains the empirical fit from 
    Leroy et al. (2024) in prep. (L24), as well as the region where \vintergatan is limited by the mass resolution of star particles when sampling the star formation model. \textit{\textbf{Takeaway:}} Star formation models based on the theory of isothermal gravo-turbulent gas relaxes the apparent contradiction seen between models of star formation and star formation scaling relations observed  in GMCs.}
    \label{fig:effphangs}
\end{figure*}

\subsection{Implications for the local \eff}

We next turn to the resulting \eff distributions for the \textit{SF-events} gas. The leftmost panel in Figure \ref{fig:effevolution} presents the evolution of the \eff and local depletion time PDFs with redshift. Given that the PN12 model is \avir-dependent, a weakly increasing virial parameter with time (solid black in Figure \ref{fig:1dh}) leads to a weak decrease of \eff. The mass-weighted mean of \eff is approximately 1\%, in agreement with the universally observed and predicted values \citep[][Semenov et al. in prep., see also \citealt{polzin24} where $\langle \epsilon_{\rm ff} \rangle \approx 1\%$ for a range of metallicities]{krumholz12,lee16,utomo18,semenov16,grisdale19}. We can rewrite the local star formation law in Equation \ref{eq:sf_schmidt} as $\rho_{\rm g}/\Dot{\rho}_{\star}=t_{\rm ff}/\epsilon_{\rm ff}$, where $t_{\rm sf} = t_{\rm ff}/\epsilon_{\rm ff}$ is the local gas depletion time. Then, in our case where the gas upon star formation in our simulation features a nearly constant \eff, $t_{\rm sf} \propto \tff$. This means that changes to the local rate of star formation are mainly driven by changes in the density PDF (as $\tff\propto \rho_{\rm g}^{-0.5}$). The rightmost panel of Figure \ref{fig:effevolution} shows the evolution of $t_{\rm sf}$. During the \textit{starburst} and \textit{LMM} epochs, both $t_{\rm sf}$ and $\tff$ (third row of Figure \ref{fig:1dh}) shifts to shorter timescales, which is coeval with the drop of global depletion time values seen in \citet{aso22}, caused by tidal compression and shocks during galaxy interactions \citep{renaud22}. In this picture, the role of increased levels of ISM turbulence is therefore \emph{not} to impact the distribution of \eff on small scales, but rather to drive the ISM density PDF towards higher densities, which in turn leads to lower global gas depletion times \citep[see also][]{kraljic14,kraljic24}. We note that the median $t_{\rm sf}$ is an order of magnitude shorter ($\sim 100 \Myr$) than the global depletion time \tdep ($\sim 1 \Gyr$) due to the fact that most gas is not star forming \citep{semenov17,polzin24}

The shape of the distributions, along with the median values (markers on the bottom of Figure \ref{fig:effevolution}), reveal a slight evolution with time: red distributions showing symbolically higher \eff medians at $z > 1$ (0.8\%) and a family of blue distributions with a factor of $\sim 3$ lower \eff after the \textit{LMM} event (0.3\%). Given the span of the \eff PDFs (0.001\%-30\%) and the more drastic evolution of $\tff$ and $\tturb$, we consider \eff to be roughly stationary. However, this change in efficiency is enough for parts of the ISM in the \textit{starburst} and \textit{LMM} periods to feature \eff above $\sim 20\%$. These star-forming regions contain the largest amount of dense turbulent gas, but as can be seen from the PDFs, they make out a small fraction of the total gas mass of the galaxy during that epoch ($\geq 2\%$ during the \textit{starburst} phase compared to $<0.5\%$ for the \textit{secular} and \textit{early} epochs). They also correspond to the shortest timescales ($< 1 \Myr$), potentially making them ideal locations for massive star cluster formation \citep{li19,li20,li22}.

In Section \ref{sec:discussion} we discuss how alternatives to PN12 could affect our conclusions.

\begin{figure*}
    \centering
    \includegraphics[width=0.85\textwidth]{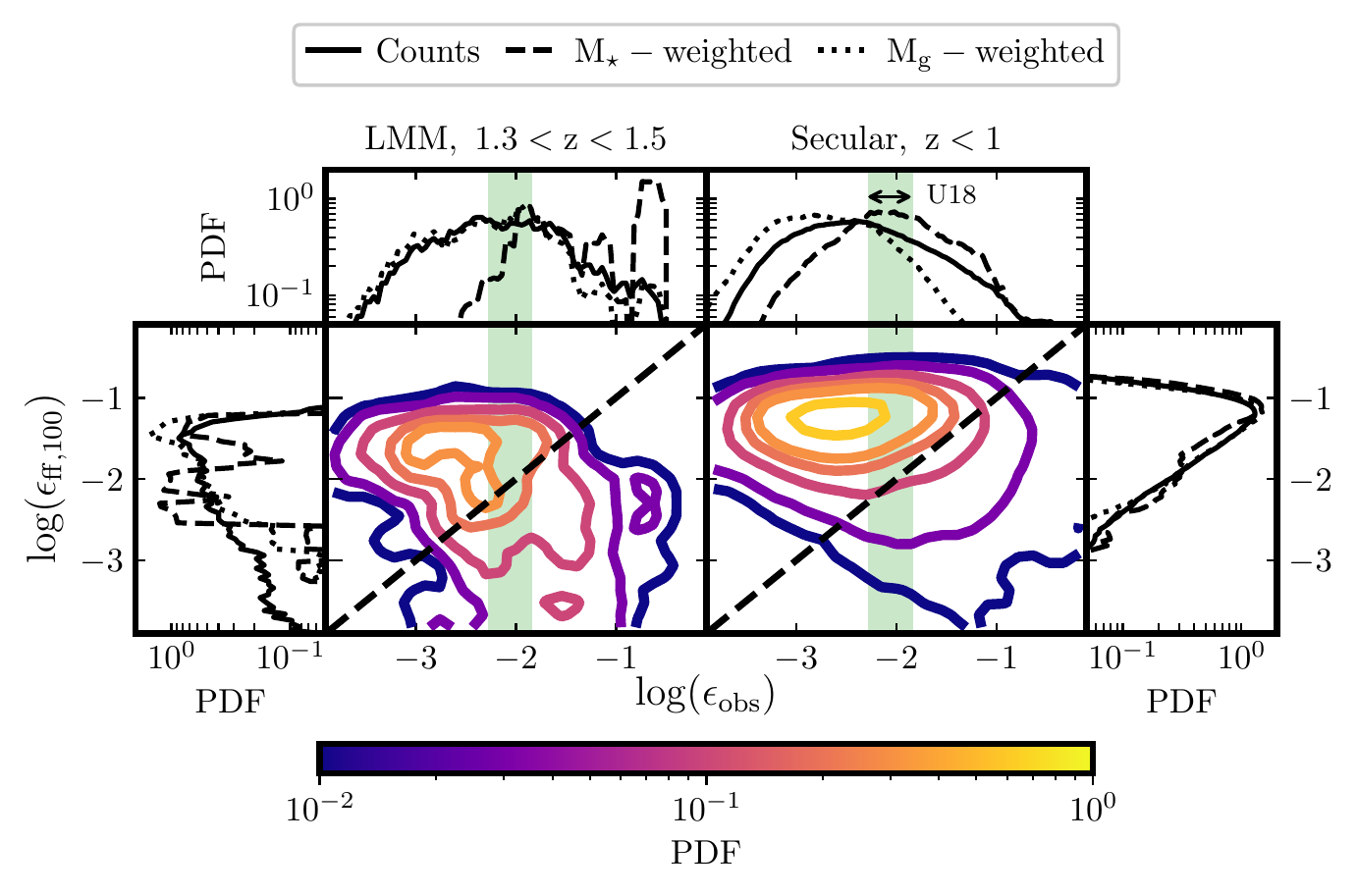}
    \caption{Comparison between the \textit{theoretically}- \citep{pn12} versus the \textit{observationally}-motivated efficiency (Equation \ref{eq:eobs}) for the \textit{LMM} and the \textit{secular} phases of \vintergatan. Both quantities are evaluated on 100 pc scales, with colour-coded contours representing the 2D histogram between \eobs and {$\epsilon_{\rm ff,100}$} indicating probability densities of 0.6, 0.3, 0.2, 0.1, 0.03, 0.01 from most yellow to most purple respectively. The corresponding 1D histograms are placed on the top and sides of the main panels, where the solid curve representing the counts is weighted by stellar mass from particles younger than 10 Myr inside the 100 pc cubes (dashed) and by the enclosed gas mass (dotted). The diagonal dashed line crossing the two main panels just shows the one-to-one correlation between \eobs and {$\epsilon_{\rm ff,100}$}. For comparison, the robust scatter of \eobs from \citet{utomo18} is highlighted as a green rectangle. \textit{\textbf{Takeaway:}} \eobs and {$\epsilon_{\rm ff,100}$} do not correlate with each other.}
    \label{fig:effvseobs}
\end{figure*}

\subsection{Inferring \eff from observations}

A large number of studies have been undertaken to constrain \eff using Milky Way GMC- and cloud-scale observations of Local Universe galaxies \citep[for reviews, see][]{Krumholz2019, schinnererleroy24} as well as in gravitationally lensed high-$z$ galaxies \citep[e.g.][]{mirka19,mirka23}. An apples-to-apples comparison with theory and simulations is not trivial \citep[see][for high resolution GMC simulations, \citealt{grisdale19} for simulations of isolated disc galaxies, \citealt{arturo21} for cosmological simulations, and \citealt{bemis23} for a more observational perspective]{grudic18,grudic19}. Observationally, \eff quantifies the ratio between recent star formation and total mass budget of GMCs, normalised by $\tff$, and probed by molecular gas and SFR tracers. Therefore, a time-lag exists between the actual trigger of star formation in a specific cloud, and the eventual signal of recent star formation (YSOs or H$\alpha$, where the latter has an associated time-lag on the order of $\sim 10 \Myr$). In contrast, in the context of theories of star formation in gravo-turbulent isothermal gas, \eff measures the amount of gas mass that is turned into stars on a free-fall timescale by integrating a log-normal density PDF above a certain density threshold. Such an \emph{instantaneous} description of \eff is derived from the current density and turbulent structure of a piece of the ISM. This is not necessarily equal to an \eff inferred from gas tracers which encodes star formation imprinted at an earlier instance of a gas cloud.

In Figure \ref{fig:effphangs} we use \vintergatan outputs in the \textit{secular} phase of evolution ($z<1$) to compute theoretically- and observationally- motivated estimates of \eff. The top row shows 2D histograms with their running medians and robust standard deviations in blue corresponding to the \textit{theoretical} star formation efficiency per free-fall time, \etheo, against the gas surface density, velocity dispersion, and virial parameter of \textit{cold+dense} gas. This time, we calculate \etheo on a grid of $100 \pc$-sized cubes using Equation \ref{eq:sf_pn12} from the velocity dispersion on these scales, and the density from the mass of \textit{cold+dense} gas within each 100 parsec cube. Note that these measurements are not necessarily the same as the on-the-fly data in Figure \ref{fig:effevolution} due to the larger sizes of the cubes with respect to the maximally resolved cells, and  the inclusion of more diffuse and less turbulent gas (Figure \ref{fig:effevolution}). We shift our analysis from 50 parsec to 100 parsec because we want to compare the predictions of the PN12 model applied on 100 pc scales with the observational \eff estimates provided by the PHANGS collaboration, most of which are also on physical sizes of $\sim 100$ parsecs \citep{rosolowsky21,sun22}. The analysis behind Figure \ref{fig:effphangs} was also performed on 50 pc scales, but no major differences were observed.

The bottom row in Figure \ref{fig:effphangs} row shows a similar analysis, only applying an \textit{observational} estimator for the star formation efficiency per free-fall time \citep{leroy17,schinnererleroy24}, hereafter \eobs:

\begin{equation}
\label{eq:eobs}
    \epsilon_{\rm obs} = \dfrac{\tff}{\tau_{\rm dep}}.
\end{equation}
Here, $\tff$ and \tdep are the local free-fall and gas depletion timescales computed from the \textit{cold+dense} gas embedding stars younger than 10 Myr. Alternatives to Equation \ref{eq:eobs} have been put forward by, e.g. \citet{lee16}\footnote{$\epsilon_{\rm obs} = \dfrac{\tff}{t_{\rm y,\star}} \dfrac{M_{\rm y,\star}}{M_{\rm g} + M_{\rm y,\star}}$, with {$M_{\rm y,\star}$} being the mass of stars younger than $t_{\rm y,\star}$.}, but our findings and main trends are not affected by the change of \eobs prescription.

It is clear from Figure \ref{fig:effphangs} that \etheo and \eobs are dramatically different measurements. Starting with the \eff-$\Sigma$ relations, \etheo increases with increasing $\Sigma$, as expected from a decreasing \avir with increasing local density. The opposite trend is found for \eobs, which stems from dense gas often probing early stages of GMCs, when only low levels of star formation have occurred and before feedback has had time to disrupt them. This yields a low mass ratio between young stars and the total mass of clouds, hence a low \eobs. The high-\eobs, low-$\Sigma$ regions of the distribution tend to represent gas-poor clouds in their final stages of evolution, where large amounts of young stars drive large estimates of \eobs \citep[][]{feldmanngnedin11,grudic18,grudic19, grisdale19}. In contrast, such diffuse environment makes the instantaneous \etheo predict very low star formation efficiencies. We note however, that the picture is much more complex than this; for instance, as seen from the \eobs-\avir relation, regions with apparent high \eobs are in fact found to be both the most bound ($\alpha_{\rm vir} < 1$) as well as unbound ($\alpha_{\rm vir} > 10$), which indicates that not all regions are necessarily undergoing late stages of star formation and feedback disruption \citep[for an example of diverse cloud evolutionary paths, see figure 3 in][]{grisdale19}.

The \eff-$\sigma$ relations predict a decreasing trend for both theoretical and observational efficiencies, yet additional efforts are required to precisely quantify whether such correlations are driven by the coupling between a multi-phase ISM and different stellar feedback mechanisms \citep[see Discussion section, but also][]{federrath15,eric24} or whether more large-scale sources of turbulence ought to be considered, e.g. shear\footnote{\textit{Cold+dense} gas in the \textit{secular} phase of \vintergatan is often located within the inner half-mass radius of the galaxy, where galactic shear has been shown to play a crucial role in shaping the local star formation properties in the ISM of these regions \citep['the Brick',][]{petkova23}.}.

To ascertain whether the star formation properties in \vintergatan at low redshifts are compatible with observations, we compare our results to resolved (100 parsec scales) results from the PHANGS collaboration. The empirical fit from Leroy et al. (2024) in prep. (hereafter L24) with its $1\sigma$ scatter is included in orange for all bottom panels in Figure \ref{fig:effphangs}. This fit was performed using cloud properties measured on pixels of 50-150 pc in size further averaged over 0.5-1.5 kpc domains and weighted by the intensity of the CO (2-1) emission line \citep[for more information see][]{sun22,sun23,schinnererleroy24}. To closely mimic such analysis, each of the \vintergatan outputs at $z < 1$ are now split in 1 kpc-sized regions. We then find which of the 100 pc cubes are located within each kpc region and re-calculate the gas mass-weighted average over the kpc domain of $\Sigma$, $\sigma$, \avir, and $\tff$. It is worth emphasizing that we do not evaluate mass-weighted averages of cell values on kpc scales, but rather use those calculated from the 100 pc-sized cubes. \eobs is still computed using Equation \ref{eq:eobs}, but we now take the gas mass-weighted free-fall time, $\langle \tff \rangle$, and divide it by the gas depletion times of each kpc domain.

Solid green lines in Figure \ref{fig:effphangs} show the running median and associated scatter through these kpc domains. We see that \vintergatan and PHANGS agree well mostly at low \eobs values, namely, in the high-$\Sigma$, high-$\sigma$ regime of the parameter space. The most striking result is appreciated in the last column of Figure \ref{fig:effphangs}. While \etheo naturally follows the PN12 star formation law with an exponential shape, the \eobs estimator recovers an almost constant star formation efficiency for virial parameters between $1 < \alpha_{\rm vir} < 10$ in accordance with PHANGS analysis. While the match between \eobs in \vintergatan and PHANGS is interesting, it remains to be seen if observed \eobs trends can be used to constrain star formation models, or whether any signature of an underlying star formation law becomes washed out due to complex overlap of diverse cloud life cycles in a single ISM patch \citep[see][]{grisdale19}. For instance, we note that the scatter in the simulation data is greater than that of the PHANGS fit. More analysis of simulations adopting a variety of star formation and feedback models is necessary to ascertain to what degree the scatter carries  information of actual cloud diversity and ISM environment, as opposed to numerical aspects. Making use of outliers to cloud distribution (e.g. cloud mass) has the potential for providing more robust insights into the physics of star forming clouds \citep{renaud24}. 

The match between \vintergatan and PHANGS at $\Sigma < 100 \Msunpcs$ and $\sigma < 100 \kms$ is poor. Part of the reason for this mismatch is inherent to \vintergatan and other cosmological simulations when sampling star formation using massive star particles, here with initial masses of $10^4\Msun$. Such high masses cannot adequately sample low levels of star formation in low density gas, hence introducing a bias. To illustrate this we added a sampling limit\footnote{The sampling limit is computed from Equation \ref{eq:eobs}, re-writing it as $\epsilon_{\rm obs} = \tff/\tau_{\rm dep} = \left(\tff/t_{\rm y,\star}\right) \cdot \left(\Sigma_{\rm y,\star}/\Sigma\right)$. Re-formulating the free-fall time as $\tff \propto \left(\Sigma/L\right)^{-0.5}$ we get that $\epsilon_{\rm obs} \propto \sqrt{L}/t_{\rm y,\star} \cdot \Sigma_{\rm y,\star}/\Sigma^{1.5}$. Here, $\Sigma_{\rm y,\star}$ corresponds to a maximally resolved star particle $\sim 10^4 \Msun$ inside a circle of radius equal to $L/2$ where $L=100 \pc$, and $t_{\rm y,\star}$ is the maximum age of young stars set to 10 Myr (Figure \ref{fig:1dh}).} with functional form $\epsilon_{\rm obs} \propto 1/\Sigma^{1.5}$. This region of the parameter space is is unreachable by our cosmological simulation and seen as a hatched grey region limited by a black dashed line in the bottom left panel of Figure \ref{fig:effphangs}. Higher resolution simulations are necessary to better probe star formation in the low density parts of the ISM, which we leave for future work.

\section{Discussion}
\label{sec:discussion}

In this paper we have studied how the density and velocity dispersion of the ISM in a Milky Way-mass galaxy evolve with cosmic time. Gravity and turbulence in gas clouds evolve conjointly, leading to an \avir distribution that is almost constant in time. Applying the PN12 recipe for star formation, we get that the local theoretical efficiency is $\langle \epsilon_{\rm ff}\rangle \approx 1\%$ computed on $20 \pc$ scales. This is in good agreement with observational efficiency estimators, e.g. in the PHANGS collaboration. In the following section we discuss in depth how mapping between the theoretical and observational efficiencies would be affected by the life cycle of clouds, merger-driven environments at high redshift, and alternative implementations of star formation.

\subsection{Mapping between \textit{theoretical} and \textit{observational} efficiencies}

Figure \ref{fig:effvseobs} explores the mapping between \eobs and \etheo over time by presenting \eobs-\etheo contours, and their corresponding 1D histograms for the \textit{LMM} event (left hand side, $1.3<z<1.5$) and the \textit{secular} epoch (right hand side, $z<1$). The distributions of counts appear as solid curves, gas mass-weighted distributions are illustrated in dotted lines, and stellar mass-weighted ones in dashed lines. Both main panels suggest a tentative anti-correlation between \eobs and \etheo.

Right hand side panels in Figure \ref{fig:effvseobs} show the \eobs-\etheo relation for \vintergatan at $z < 1$. Surprisingly, an order of magnitude difference exists between a median \etheo of $\sim 4\%$ (rightmost panel) and a median \eobs of $\sim 0.4\%$ (top right panel) regardless of the distributions being stellar or gas mass-weighted. This suggests that: $i)$ from a PN12 point of view the ISM is considered to be more efficiently star-forming on 100 pc scales and free-fall timescales of $\sim 3.5 \Myr$ than what \eobs implies; $ii)$ the observational estimator \eobs describes a less efficiently star forming ISM with gas depletion times of $\sim 300 \Myr$. From Figure \ref{fig:effphangs} we concluded that star-forming clouds in their initial stages of evolution are dense and turbulent, with gas masses larger than that of stars younger than 10 Myr, appearing inefficient when traced by observations yet efficient as predicted by theory. Contour lines in Figure \ref{fig:effvseobs} do show a tendency towards the low-\etheo, high-\eobs corner of the panel indicating such cloud evolution. Nevertheless, some of these early clouds are found in the central regions of \vintergatan and have the highest SFRs, so even after the onset of the first SNe, it is not clear whether feedback will be able to couple to such large cold and dense gas reservoirs to impact \avir \citep[][see \citealt{hopkins13b} for feedback-regulated star formation, where SNe feedback weakly couples to cold and dense gas clouds with high HCN/CO compared to pre-SNe feedback, \citealt{ohlin19} for tests of SNe explosions in turbulent ISM boxes, and \citealt{zakardjian23} for spatial offsets between HCN and CO peaks]{querejeta19}.

During the \textit{LMM} phase, \vintergatan is undergoing a starburst. Panels on the left hand side of Figure \ref{fig:effvseobs} illustrate how the \eobs-\etheo trend remains visible in a high-$z$ major merger, almost orthogonal to the one-to-one relation (black dashed line). If we look at the distribution of counts or gas mass-weighted \eff values (solid and dotted respectively), there is a minor tendency towards lower \etheo and higher \eobs compared to the \textit{secular} phase, with median $\epsilon_{\rm ff,100} \sim \epsilon_{\rm obs} \sim 1\%$ (leftmost and top left panels). In turn, the stellar mass-weighted \eobs distributions are significantly shifted towards high efficiencies of $\sim 10\%$ (dashed lines). This is because shorter local gas depletion times of $\sim 100 \Myr$ (Figure \ref{fig:1dh}), imply that there is an enhancement of star formation on shorter timescales when \vintergatan is observed as a starburst galaxy rather than a main-sequence galaxy. On 100 pc scales, this translates into larger stellar-to-gas mass ratios, which skews the stellar-mass weighted \eobs distributions. Such efficiency values have been reported from YSOs in dense clumps in the Milky Way \citep{heyer16}, dense gas tracers like HCN and CS from extra-galactic sources \citep{wu10}, and from dense star forming regions in starbursting high-redshift lensed galaxies  \citep{mirka19,mirka23}.

Interestingly, the {$\epsilon_{\rm ff,100}$} distribution indicates less efficient star formation during the \textit{LMM} (leftmost panel) than in the \textit{secular} phase (rightmost panel) due to enhanced \avir. This is at odds with Figure \ref{fig:effevolution}, which characterises \vintergatan as a marginally less efficient galaxy after the \textit{LMM} as gas is no longer in high density states. Even though both {$\epsilon_{\rm ff,100}$} and \eff are computed using Equation \ref{eq:sf_pn12}, differences arise due to the different scales in which these parameters are computed, but most importantly due to the fact that \textit{SF-events} and \textit{cold+dense} describe inherently different states of gas. Namely, \textit{SF-events} only include gas that has eventually formed a star, with moderately larger velocity dispersion values yet substantially denser gas. This makes up for an \eff evolution that follows changes in the gas density PDF.

Our results indicate that the multi-scale aspect of deriving star formation efficiencies as well as the use of different SFR and molecular gas tracers ought to be considered. These will weigh \eff distributions differently and imply various degrees of feedback coupling to the surrounding ISM that must ultimately be taken into account when comparing star formation models to observational estimators \citep[][and references therein]{utomo18,Krumholz2019,schinnererleroy24}.

\begin{figure}
    \centering
    \includegraphics[width=\columnwidth]{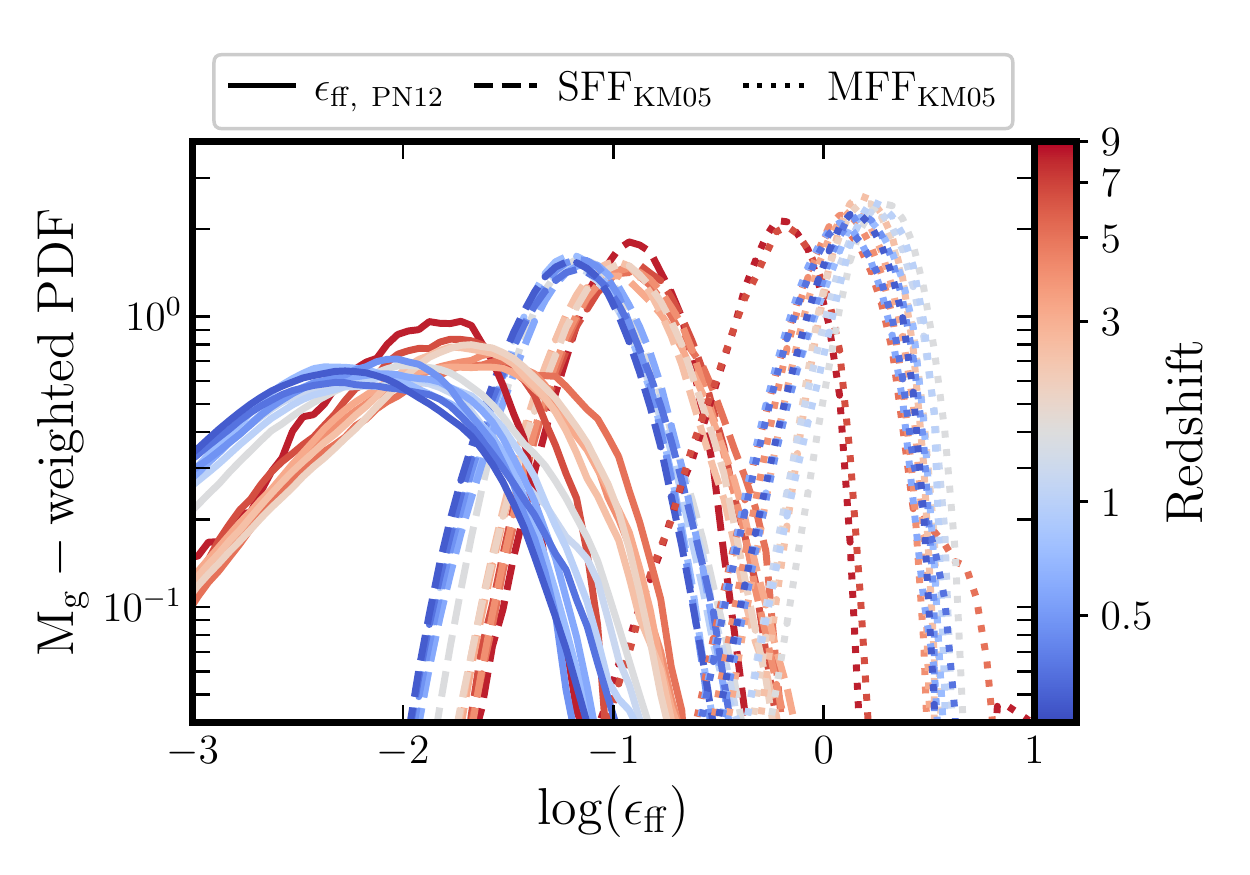}
    \caption{Comparison among three \eff models and the evolution of their gas mass-weighted PDFs with redshift. While the solid lines outline the on-the-fly \eff PDFs in \vintergatan, the same {$n$} and {$\sigma$} from the \textit{SF-events} gas is used as input in the other two models. No additional simulations have been run with different \eff models. From left to right we have: {$\epsilon_{\rm ff, \ PN12}$} solid lines represent the PN12 prescription, SFF{$_{\rm KM05}$} dashed distributions stand for the single free-fall model adopted by \citet{fk12} from \citet{km05}, and the MFF{$_{\rm KM05}$} in dotted PDFs correspond to the multi free-fall adaptation of the SFF{$_{\rm KM05}$} based on \citet{hc11}. \textit{\textbf{Takeaway:}} The same on-the-fly density and velocity dispersion values are used as inputs for all models, so it is just a prove of concept that, \textit{a priori}, different models result in different \eff distributions.}
    \label{fig:effmodels}
\end{figure}

\subsection{Alternative star formation efficiency models}
In this section we discuss how our findings could be affected when changing sub-grid recipes for star formation, and how can that modify the structure of the ISM. In order to illustrate the sensitivity of \eff to different star formation recipes, Figure \ref{fig:effmodels} illustrates three \eff models for a fixed gas distribution, i.e. for the same cell density and velocity dispersion input without directly re-running \vintergatan for a self-consistent ISM:

\begin{itemize}
    \item \textit{\textbf{PN12:}} star formation prescription used in the \vintergatan simulation, presented in Equations \ref{eq:sf_schmidt} and \ref{eq:sf_pn12} (solid lines in Figure \ref{fig:effmodels}). This model has been calibrated on high resolution simulations of magnetised ISM boxes \citep{pn12} that cover a range in virial parameters of {$0.3 \geq \alpha_{\rm vir} \geq 13$}. The main take away from the PN12 model is that \eff exponentially decreases with increasing $\tff/\tturb$, and changes by less than a factor of two with magnetic field strength for typical star forming regions.
    \item \textit{\textbf{SFF$_{\rm KM05}$:}} single free-fall models (SFF) define a family equations that explicitly integrate the log-normal distribution of gas densities to calculate the fraction of dense gas above a critical density that undergoes star formation on a \textit{fixed} free-fall timescale\footnote{$\epsilon_{\rm ff} = \dfrac{\epsilon_{w}}{\phi_{t}} \int^{\infty}_{s_{\rm crit}} \dfrac{t_{\rm ff}\left(\rho_{0}\right)}{t_{\rm ff}\left(\rho\right)} \dfrac{\rho}{\rho_{0}} p\left(s\right) ds$, with $p\left(s\right)$ being the log-normal distribution of $s=\ln{\left( \rho/\rho_{0}\right)}$, $s_{\rm crit}$ the critical density equal to $(\pi^{2}/5) \phi_{x}^{2} \times \alpha_{\rm vir} \mathcal{M}^{2}$ for the hydrodynamics-only case in both the SFF{$_{\rm KM05}$} and MFF{$_{\rm KM05}$} models \citep{fk12}, and $\tff \left(\rho\right)$ is the free-fall timescale set to $\tff \left(\rho_{0}\right)$ in SFF{$_{\rm KM05}$} but kept inside the integral for MFF{$_{\rm KM05}$}. Both $\phi_{t}$ and $\phi_{x}$ are calibration factors set to {$1/\phi_{t} = 3$} and {$\phi_{x} = 0.12$} for SFF{$_{\rm KM05}$} and {$1/\phi_{t} = 0.49$} and {$\phi_{x} = 0.19$} for MFF{$_{\rm KM05}$} in Figure \ref{fig:effmodels}. The turbulent forcing parameter is {$b_{\rm turb} = 0.38$}, representing a mix of solenoidal and compressive modes of turbulence}. Dashed curves in Figure \ref{fig:effmodels} outline the star formation model based on equation 20 in \citet{km05}, and further adapted by \citet{fk12} (see their table 1). In this model, the free-fall time is fixed to that of the average density of the cloud.
    \item \textit{\textbf{MFF$_{\rm KM05}$:}} multi-free-fall models (MFF) as dotted curves in Figure \ref{fig:effmodels} describe a family of star formation equation similar to SFF ones only that they allow the critical density threshold to vary given that gravitationally bound structures of different densities will collapse at different $\tff$ \citep[see][and table 1 in \citealt{fk12}]{hc11}.
\end{itemize}

The main takeaway from Figure \ref{fig:effmodels} is that, the choice of \eff prescription shifts the emerging $\langle \epsilon_{\rm ff} \rangle$. Both SFF{$_{\rm KM05}$} and MFF{$_{\rm KM05}$} render $\langle \epsilon_{\rm ff} \rangle$ of 10\% and 100\% respectively. Nevertheless, we insist on the fact that a rigorous comparison would involve explicitly resolving the star formation-feedback cycle, leading to potentially different observables such as the KS relation, global \tdep evolution across cosmic time, etc.

From the \avir distributions in Figure \ref{fig:1dh} we see that at large fraction of the star-forming ISM has virial parameters that exceed those of the MHD ISM boxes used for the calibration of PN12, with Mach numbers approaching  {$\sim 100$}. These are extreme conditions, where PN12 is extrapolated beyond its \avir limits towards values below 1 percent. Both SFF{$_{\rm KM05}$} and MFF{$_{\rm KM05}$} are \avir and $\mach$ dependent, meaning that for the same cell density input, additional dependencies on its velocity dispersion will shift \eff distributions towards more efficient values. That is particularly the case for the MFF{$_{\rm KM05}$} model, with an extra factor {$\exp{\left[\left(3/8\right) \sigma_{s}^2\right]}$}, where {$\sigma_{s} = \ln{\left(1+b^2\mathcal{M}^2\right)}$}. In the future we will evolve self-consistently these models using \vintergatan physics for strict comparison.

The role of turbulence is critical in regulating star formation \citep{renaud12,kraljic24}, but the details in which it does so still remain unsettled. The reason is that since the dynamical range of the turbulent cascade of energy is not fully resolved in simulations, sub-grid recipes are developed to compensate for the effect of the small-scale turbulent eddies. In a recent study, \citet{brucy24a,brucy24b} emphasise that the density PDF at high Mach numbers can even depart from the often assumed log-normal PDF shapes \citep{castaing96,hopkins13a}. These are promising simulations of high resolution isothermal ISM boxes that will soon be tested in the context of isolated disc and cosmological simulations. \citet{semenov16} implemented the sub-grid scale turbulent model \citep[SGS,][]{schmidtfederrath11,schmidt14} where the production and dissipation of unresolved turbulent energy is followed explicitly and used to calculate \avir and \eff. Their isolated galaxy simulation well reproduces observed \eff distributions and the KS relation, but lacks the cosmological context. In fact, their work will be extended to cosmological zoom-in simulations in Semenov et al. in prep., where a near-universal \eff of $\sim 1\%$ is also found when modelling the local variations of \eff from the explicitly followed turbulent energy on unresolved scales. \citet{kretschmer20} and \citet{arturo21} recently ran cosmological simulations using MFF in combination with SGS models and compare it to a constant \eff of 1 percent. They noticed how \eff is much more sensitive to the cosmological environment than what we see in \vintergatan. They conclude that it is the combination of star formation and feedback that regulates the structure of the ISM and eventually the emerging GMC and galactic-scale star formation scaling relations.

One of the main results from the FIRE collaboration \citep{fauchergiguere13,hopkins14,orr18,hopkins18,feldmann23} is that feedback is central to regulating star formation. At sufficiently high resolution, gas cools down, becomes increasingly dense, and forms stars until feedback disperses the cloud \citep[see discussion in][]{hopkins12}. \cite{hopkins11} argue that a lower value of \eff simply allows for gravitational collapse to proceed for longer, eventually leading to a short enough $t_{\rm sf}=t_{\rm ff}/$\eff, allowing for fast local star formation and clustered feedback, which in turn leads to cloud disruption. However, while changes to \eff, as well as the criteria for star formation, can lead to little-to-no impact on integrated quantities on galactic scales, e.g. global SFRs, the emerging  structure of the ISM can vary significantly \citep[e.g.][]{hopkins13c}.  

When feedback is not as efficient, it is \eff that regulates star formation \citep{semenov17}. The \eff can even lead to diverging global properties such as SFRs \citep{benincasa16} or the normalisation of the KS relation \citep[factor of a few,][]{agertz13}. We ultimately want to understand the underlying agent that determines how much gas mass and on which timescales does it pile up at the threshold of star formation. Targeting GMC properties like those in Figure \ref{fig:effphangs} stand as stringent probes to distinguish star formation models, but as concluded from our results and along with previous studies, \eff alone does not fully inform on the SFR in the cell. It is the star formation-feedback cycle that eventually regulates star formation and that will set the observed \eff \citep[see discussion in][]{federrath15}.

\section{Conclusions}
\label{sec:conclusions}

In this paper we analyse the density and velocity dispersion of the ISM and its implications on the local star formation efficiency per free-fall time in \vintergatan, a high resolution cosmological zoom-in simulation of a Milky Way-type galaxy \citep{vinter1}. We follow the evolution of their PDFs in the context of JWST observations \citep{heintz22,aravena23}, lensed galaxies and mergers around the cosmic peak of star formation \citep{mirka19,mirka23,rizzo23,romanolveira23}, and GMC observables at {$z \approx 0$} from the PHANGS collaboration \citep[][Leroy et al. 2024 in prep.]{sun18,sun20b,rosolowsky21,schinnererleroy24}. Our findings can be summarised in four points:

\begin{itemize}
    \item \textbf{The density and turbulent structure of the star forming ISM on 50 pc scales reacts to its extra-galactic environment} (Figure \ref{fig:threephaseism}). Three cosmic epochs are identified: $i)$ \textit{early on} ($5<z<9$), where the gas-rich proto-galaxy is being assembled mostly from in-situ star formation and minor contributions from neighbouring galaxies that sustain moderate velocity dispersions levels \citep[$\sigma < 30 \kms$,][]{pope23,posses23}; $ii)$ a \textit{starburst} epoch ($1<z<5$) that ends with the \textit{last major merger} event ($1.3<z<1.5$), where the galaxy builds a rotationally supported disc and experiences merger-driven starbursts that compress the ISM to $\sigma \approx 100 \kms$ \citep{herreracamus22,parlanti23} with global gas depletion times of $\sim 0.1 \Gyr$ due to an excess of gas at high densities; $iii)$ a final \textit{secular} phase where \vintergatan settles into an extended main sequence galaxy where $\sigma \approx 10 \kms$ \citep{tamburro09}.

    \item \textbf{The density and velocity dispersion distributions on 50 pc scales evolve with cosmic time, but conspire to yield virial parameter distributions that are nearly constant} (Figure \ref{fig:1dh}). The resulting ISM at low redshift has virial parameter distributions within $1 < \alpha_{\rm vir} < 10$, values that are in agreement with observations \citep{leroy15,mivilledeschenes17,mirka19,rosolowsky21}. The more exposed the ISM is to recent star formation, and therefore affected by the subsequent stellar feedback, the denser and more turbulent it is. The presence of mergers, or absence thereof, pushes the ISM to such extreme conditions, nevertheless, their exact role in shaping \avir distributions is still unclear.

    \item \textbf{The emerging \eff distribution spans several orders of magnitude, from {$0.001\%$} up to {$30\%$}, but averages to a roughly constant {$\langle \epsilon_{\rm ff} \rangle \approx 1\%$} across cosmic time} (Figure \ref{fig:effevolution}) in line with observations \citep[][]{lee16,utomo18,Krumholz2019} and recent computationally-driven studies \citep[][]{semenov16,kretschmer20,arturo21,polzin24}. The model adopted in \vintergatan is based on \citet{pn12} and is exclusively \avir-dependent. Thus, for a constant \eff the local gas depletion time scales as {$t_{\rm sf} \sim t_{\rm ff}/\epsilon_{\rm ff} \sim t_{\rm ff} \propto \rho_{\rm g}^{-0.5}$}. As such, changes in the gas density PDFs become the main controlling factor in setting the local rate of star formation. The role of turbulence is to instead establish the density PDF in the first place \citep{kraljic14,renaud14,aso22,kraljic24}, rather than inducing large changes to \eff.
    
    \item \textbf{\textit{Theoretical}, more \textit{instantaneous} star formation efficiencies are inherently different to their \textit{observational} estimators as they represent GMCs in different physical states} (Figure \ref{fig:effphangs}). Different star-forming gas tracers (CO, HCN, CS, etc.) probe a range of densities and temperatures. Tracers of recent star formation (YSOs, H${\alpha}$, UV, etc.) also provide information about the star-forming cloud, but it must be \textit{integrated} over the lifetime of the tracer employed \citep{grudic19}. This often hinders us from accessing the local instantaneous conditions for star formation. Figure \ref{fig:effvseobs} hints towards a possible anti-correlation between the theoretical and observational \eff, proposing that clouds in their initial stages of evolution begin with high \etheo and low \eobs, and end their lives with low \etheo and high \eobs. Our conclusions cast light onto seemingly contradictory results between star formation models and observations of star-forming GMC \citep{schruba19}. This behaviour is consistent with PHANGS observations \citep[][Leroy et al. 2024 in prep]{leroy17,schinnererleroy24}.
\end{itemize}

Linking local star formation parameters from ISM properties is highly non-linear and complex and has to rely on well calibrated sub-grid models \citep{nobles24}. These are rapidly improving with time \citep[see][who include magnetic fields and \citealt{polzin24} who add a metallicity dependency]{girma24}, but still do not consider the effects of the larger-scale environment (e.g. shear, compression, etc.). 

In summary, star formation models, while degenerate, can be singled out when comparing to observed ISM properties. Moreover, information can be obtained from the scatter in GMC scaling relations, but how exactly it can help constrain star formation models is yet to be pinned down. In our work we stress the need for understanding the coupling among star formation, feedback, and a multi-phase ISM. Thus, a change in star formation models must go hand-in-hand with a change in the feedback recipe. Further observational data will be soon available in the advent of upcoming JWST and ALMA observations, and the further development of dedicated surveys like PHANGS and MANGA, generating multi-scale multi-wavelength data products of cloud properties.

\section*{Acknowledgements}

We are deeply grateful to Adam K. Leroy for providing the empirical data from the PHANGS Collaboration and to Marta Reina Campos for insightful discussions. OA and ASO acknowledge support from the Knut and Alice Wallenberg Foundation, the Swedish Research Council (grant 2019-04659), and the Swedish National Space Agency (SNSA Dnr 2023-00164). FR acknowledges support provided by the University of Strasbourg Institute for Advanced Study (USIAS), within the French national program Investment for the Future (Excellence Initiative) IdEx-Unistra. Support for VS was provided by Harvard University through the Institute for Theory and Computation Fellowship. 

\section*{Data Availability}
 
The data underlying this article will be shared on reasonable request to the corresponding author.

\bibliographystyle{mnras}
\bibliography{eff}

\begin{thebibliography}{}
\makeatletter
\relax
\def\mn@urlcharsother{\let\do\@makeother \do\$\do\&\do\#\do\^\do\_\do\%\do\~}
\def\mn@doi{\begingroup\mn@urlcharsother \@ifnextchar [ {\mn@doi@}
  {\mn@doi@[]}}
\def\mn@doi@[#1]#2{\def\@tempa{#1}\ifx\@tempa\@empty \href
  {http://dx.doi.org/#2} {doi:#2}\else \href {http://dx.doi.org/#2} {#1}\fi
  \endgroup}
\def\mn@eprint#1#2{\mn@eprint@#1:#2::\@nil}
\def\mn@eprint@arXiv#1{\href {http://arxiv.org/abs/#1} {{\tt arXiv:#1}}}
\def\mn@eprint@dblp#1{\href {http://dblp.uni-trier.de/rec/bibtex/#1.xml}
  {dblp:#1}}
\def\mn@eprint@#1:#2:#3:#4\@nil{\def\@tempa {#1}\def\@tempb {#2}\def\@tempc
  {#3}\ifx \@tempc \@empty \let \@tempc \@tempb \let \@tempb \@tempa \fi \ifx
  \@tempb \@empty \def\@tempb {arXiv}\fi \@ifundefined
  {mn@eprint@\@tempb}{\@tempb:\@tempc}{\expandafter \expandafter \csname
  mn@eprint@\@tempb\endcsname \expandafter{\@tempc}}}

\bibitem[\protect\citeauthoryear{{Agertz} \& {Kravtsov}}{{Agertz} \&
  {Kravtsov}}{2015}]{agertzkravtsov15}
{Agertz} O.,  {Kravtsov} A.~V.,  2015, \mn@doi [\apj]
  {10.1088/0004-637X/804/1/18}, \href
  {https://ui.adsabs.harvard.edu/abs/2015ApJ...804...18A} {804, 18}

\bibitem[\protect\citeauthoryear{{Agertz}, {Teyssier}  \& {Moore}}{{Agertz}
  et~al.}{2011}]{agertz11}
{Agertz} O.,  {Teyssier} R.,   {Moore} B.,  2011, \mn@doi [\mnras]
  {10.1111/j.1365-2966.2010.17530.x}, \href
  {https://ui.adsabs.harvard.edu/abs/2011MNRAS.410.1391A} {410, 1391}

\bibitem[\protect\citeauthoryear{{Agertz}, {Kravtsov}, {Leitner}  \&
  {Gnedin}}{{Agertz} et~al.}{2013}]{agertz13}
{Agertz} O.,  {Kravtsov} A.~V.,  {Leitner} S.~N.,   {Gnedin} N.~Y.,  2013,
  \mn@doi [\apj] {10.1088/0004-637X/770/1/25}, \href
  {https://ui.adsabs.harvard.edu/abs/2013ApJ...770...25A} {770, 25}

\bibitem[\protect\citeauthoryear{{Agertz} et~al.,}{{Agertz}
  et~al.}{2020}]{agertz20}
{Agertz} O.,  et~al., 2020, \mn@doi [\mnras] {10.1093/mnras/stz3053}, \href
  {https://ui.adsabs.harvard.edu/abs/2020MNRAS.491.1656A} {491, 1656}

\bibitem[\protect\citeauthoryear{{Agertz} et~al.,}{{Agertz}
  et~al.}{2021}]{vinter1}
{Agertz} O.,  et~al., 2021, \mn@doi [\mnras] {10.1093/mnras/stab322}, \href
  {https://ui.adsabs.harvard.edu/abs/2021MNRAS.503.5826A} {503, 5826}

\bibitem[\protect\citeauthoryear{{Andersson}, {Mac Low}, {Agertz}, {Renaud}  \&
  {Li}}{{Andersson} et~al.}{2024}]{eric24}
{Andersson} E.~P.,  {Mac Low} M.-M.,  {Agertz} O.,  {Renaud} F.,   {Li} H.,
  2024, \mn@doi [\aap] {10.1051/0004-6361/202347792}, \href
  {https://ui.adsabs.harvard.edu/abs/2024A&A...681A..28A} {681, A28}

\bibitem[\protect\citeauthoryear{{Aravena} et~al.,}{{Aravena}
  et~al.}{2023}]{aravena23}
{Aravena} M.,  et~al., 2023, \mn@doi [arXiv e-prints]
  {10.48550/arXiv.2309.15948}, \href
  {https://ui.adsabs.harvard.edu/abs/2023arXiv230915948A} {p. arXiv:2309.15948}

\bibitem[\protect\citeauthoryear{{Aubert} \& {Teyssier}}{{Aubert} \&
  {Teyssier}}{2010}]{auberteyssier10}
{Aubert} D.,  {Teyssier} R.,  2010, \mn@doi [\apj]
  {10.1088/0004-637X/724/1/244}, \href
  {https://ui.adsabs.harvard.edu/abs/2010ApJ...724..244A} {724, 244}

\bibitem[\protect\citeauthoryear{{Bemis} \& {Wilson}}{{Bemis} \&
  {Wilson}}{2023}]{bemis23}
{Bemis} A.~R.,  {Wilson} C.~D.,  2023, \mn@doi [\apj]
  {10.3847/1538-4357/acb352}, \href
  {https://ui.adsabs.harvard.edu/abs/2023ApJ...945...42B} {945, 42}

\bibitem[\protect\citeauthoryear{{Benincasa}, {Wadsley}, {Couchman}  \&
  {Keller}}{{Benincasa} et~al.}{2016}]{benincasa16}
{Benincasa} S.~M.,  {Wadsley} J.,  {Couchman} H.~M.~P.,   {Keller} B.~W.,
  2016, \mn@doi [\mnras] {10.1093/mnras/stw1741}, \href
  {https://ui.adsabs.harvard.edu/abs/2016MNRAS.462.3053B} {462, 3053}

\bibitem[\protect\citeauthoryear{{Bertoldi} \& {McKee}}{{Bertoldi} \&
  {McKee}}{1992}]{bertoldimckee92}
{Bertoldi} F.,  {McKee} C.~F.,  1992, \mn@doi [\apj] {10.1086/171638}, \href
  {https://ui.adsabs.harvard.edu/abs/1992ApJ...395..140B} {395, 140}

\bibitem[\protect\citeauthoryear{{Bolatto}, {Leroy}, {Rosolowsky}, {Walter}  \&
  {Blitz}}{{Bolatto} et~al.}{2008}]{bolatto08}
{Bolatto} A.~D.,  {Leroy} A.~K.,  {Rosolowsky} E.,  {Walter} F.,   {Blitz} L.,
  2008, \mn@doi [\apj] {10.1086/591513}, \href
  {https://ui.adsabs.harvard.edu/abs/2008ApJ...686..948B} {686, 948}

\bibitem[\protect\citeauthoryear{{Brinchmann}, {Charlot}, {White}, {Tremonti},
  {Kauffmann}, {Heckman}  \& {Brinkmann}}{{Brinchmann}
  et~al.}{2004}]{brinchmann04}
{Brinchmann} J.,  {Charlot} S.,  {White} S.~D.~M.,  {Tremonti} C.,  {Kauffmann}
  G.,  {Heckman} T.,   {Brinkmann} J.,  2004, \mn@doi [\mnras]
  {10.1111/j.1365-2966.2004.07881.x}, \href
  {https://ui.adsabs.harvard.edu/abs/2004MNRAS.351.1151B} {351, 1151}

\bibitem[\protect\citeauthoryear{{Brucy}, {Hennebelle}  \& {Colman}}{{Brucy}
  et~al.}{2024a}]{brucy24a}
{Brucy} N.,  {Hennebelle} P.,   {Colman} T.,  2024a, \mn@doi [arXiv e-prints]
  {10.48550/arXiv.2404.17368}, \href
  {https://ui.adsabs.harvard.edu/abs/2024arXiv240417374B} {p. arXiv:2404.17368}

\bibitem[\protect\citeauthoryear{{Brucy}, {Hennebelle}, {Colman}, {Klessen}  \&
  {Le Yhuelic}}{{Brucy} et~al.}{2024b}]{brucy24b}
{Brucy} N.,  {Hennebelle} P.,  {Colman} T.,  {Klessen} R.~S.,   {Le Yhuelic}
  C.,  2024b, \mn@doi [arXiv e-prints] {10.48550/arXiv.2404.17374}, \href
  {https://ui.adsabs.harvard.edu/abs/2024arXiv240417374B} {p. arXiv:2404.17374}

\bibitem[\protect\citeauthoryear{{Burkhart}}{{Burkhart}}{2018}]{burkhart18}
{Burkhart} B.,  2018, \mn@doi [\apj] {10.3847/1538-4357/aad002}, \href
  {https://ui.adsabs.harvard.edu/abs/2018ApJ...863..118B} {863, 118}

\bibitem[\protect\citeauthoryear{{Castaing}}{{Castaing}}{1996}]{castaing96}
{Castaing} B.,  1996, \mn@doi [Journal de Physique II] {10.1051/jp2:1996172},
  \href {https://ui.adsabs.harvard.edu/abs/1996JPhy2...6..105C} {6, 105}

\bibitem[\protect\citeauthoryear{{Chabrier}}{{Chabrier}}{2003}]{chabrier03}
{Chabrier} G.,  2003, \mn@doi [\pasp] {10.1086/376392}, \href
  {https://ui.adsabs.harvard.edu/abs/2003PASP..115..763C} {115, 763}

\bibitem[\protect\citeauthoryear{{Ciesla} et~al.,}{{Ciesla}
  et~al.}{2023}]{ciesla23}
{Ciesla} L.,  et~al., 2023, \mn@doi [\aap] {10.1051/0004-6361/202245376}, \href
  {https://ui.adsabs.harvard.edu/abs/2023A&A...672A.191C} {672, A191}

\bibitem[\protect\citeauthoryear{{Courty} \& {Alimi}}{{Courty} \&
  {Alimi}}{2004}]{courtyalimi04}
{Courty} S.,  {Alimi} J.~M.,  2004, \mn@doi [\aap]
  {10.1051/0004-6361:20031736}, \href
  {https://ui.adsabs.harvard.edu/abs/2004A&A...416..875C} {416, 875}

\bibitem[\protect\citeauthoryear{{Daddi} et~al.,}{{Daddi}
  et~al.}{2010}]{daddi10}
{Daddi} E.,  et~al., 2010, \mn@doi [\apjl] {10.1088/2041-8205/714/1/L118},
  \href {https://ui.adsabs.harvard.edu/abs/2010ApJ...714L.118D} {714, L118}

\bibitem[\protect\citeauthoryear{{Dessauges-Zavadsky}
  et~al.,}{{Dessauges-Zavadsky} et~al.}{2019}]{mirka19}
{Dessauges-Zavadsky} M.,  et~al., 2019, \mn@doi [Nature Astronomy]
  {10.1038/s41550-019-0874-0}, \href
  {https://ui.adsabs.harvard.edu/abs/2019NatAs...3.1115D} {3, 1115}

\bibitem[\protect\citeauthoryear{{Dessauges-Zavadsky}
  et~al.,}{{Dessauges-Zavadsky} et~al.}{2023}]{mirka23}
{Dessauges-Zavadsky} M.,  et~al., 2023, \mn@doi [\mnras]
  {10.1093/mnras/stad113}, \href
  {https://ui.adsabs.harvard.edu/abs/2023MNRAS.519.6222D} {519, 6222}

\bibitem[\protect\citeauthoryear{{Duncan} et~al.,}{{Duncan}
  et~al.}{2019}]{duncan19}
{Duncan} K.,  et~al., 2019, \mn@doi [\apj] {10.3847/1538-4357/ab148a}, \href
  {https://ui.adsabs.harvard.edu/abs/2019ApJ...876..110D} {876, 110}

\bibitem[\protect\citeauthoryear{{Eibensteiner} et~al.,}{{Eibensteiner}
  et~al.}{2023}]{eibensteiner23}
{Eibensteiner} C.,  et~al., 2023, \mn@doi [\aap] {10.1051/0004-6361/202245290},
  \href {https://ui.adsabs.harvard.edu/abs/2023A&A...675A..37E} {675, A37}

\bibitem[\protect\citeauthoryear{{Ejdetj{\"a}rn}, {Agertz}, {{\"O}stlin},
  {Renaud}  \& {Romeo}}{{Ejdetj{\"a}rn} et~al.}{2022}]{timmy22}
{Ejdetj{\"a}rn} T.,  {Agertz} O.,  {{\"O}stlin} G.,  {Renaud} F.,   {Romeo}
  A.~B.,  2022, \mn@doi [\mnras] {10.1093/mnras/stac1414}, \href
  {https://ui.adsabs.harvard.edu/abs/2022MNRAS.514..480E} {514, 480}

\bibitem[\protect\citeauthoryear{{Elmegreen}}{{Elmegreen}}{1989}]{Elmegreen1989}
{Elmegreen} B.~G.,  1989, \mn@doi [\apj] {10.1086/167192}, \href
  {https://ui.adsabs.harvard.edu/abs/1989ApJ...338..178E} {338, 178}

\bibitem[\protect\citeauthoryear{{Faucher-Gigu{\`e}re}, {Quataert}  \&
  {Hopkins}}{{Faucher-Gigu{\`e}re} et~al.}{2013}]{fauchergiguere13}
{Faucher-Gigu{\`e}re} C.-A.,  {Quataert} E.,   {Hopkins} P.~F.,  2013, \mn@doi
  [\mnras] {10.1093/mnras/stt866}, \href
  {https://ui.adsabs.harvard.edu/abs/2013MNRAS.433.1970F} {433, 1970}

\bibitem[\protect\citeauthoryear{{Federrath}}{{Federrath}}{2015}]{federrath15}
{Federrath} C.,  2015, \mn@doi [\mnras] {10.1093/mnras/stv941}, \href
  {https://ui.adsabs.harvard.edu/abs/2015MNRAS.450.4035F} {450, 4035}

\bibitem[\protect\citeauthoryear{{Federrath} \& {Klessen}}{{Federrath} \&
  {Klessen}}{2012}]{fk12}
{Federrath} C.,  {Klessen} R.~S.,  2012, \mn@doi [\apj]
  {10.1088/0004-637X/761/2/156}, \href
  {https://ui.adsabs.harvard.edu/abs/2012ApJ...761..156F} {761, 156}

\bibitem[\protect\citeauthoryear{{Federrath} \& {Klessen}}{{Federrath} \&
  {Klessen}}{2013}]{federrath13}
{Federrath} C.,  {Klessen} R.~S.,  2013, \mn@doi [\apj]
  {10.1088/0004-637X/763/1/51}, \href
  {https://ui.adsabs.harvard.edu/abs/2013ApJ...763...51F} {763, 51}

\bibitem[\protect\citeauthoryear{{Federrath}, {Roman-Duval}, {Klessen},
  {Schmidt}  \& {Mac Low}}{{Federrath} et~al.}{2010}]{federrath10}
{Federrath} C.,  {Roman-Duval} J.,  {Klessen} R.~S.,  {Schmidt} W.,   {Mac Low}
  M.~M.,  2010, \mn@doi [\aap] {10.1051/0004-6361/200912437}, \href
  {https://ui.adsabs.harvard.edu/abs/2010A&A...512A..81F} {512, A81}

\bibitem[\protect\citeauthoryear{{Feldmann} \& {Gnedin}}{{Feldmann} \&
  {Gnedin}}{2011}]{feldmanngnedin11}
{Feldmann} R.,  {Gnedin} N.~Y.,  2011, \mn@doi [\apjl]
  {10.1088/2041-8205/727/1/L12}, \href
  {https://ui.adsabs.harvard.edu/abs/2011ApJ...727L..12F} {727, L12}

\bibitem[\protect\citeauthoryear{{Feldmann} et~al.,}{{Feldmann}
  et~al.}{2023}]{feldmann23}
{Feldmann} R.,  et~al., 2023, \mn@doi [\mnras] {10.1093/mnras/stad1205}, \href
  {https://ui.adsabs.harvard.edu/abs/2023MNRAS.522.3831F} {522, 3831}

\bibitem[\protect\citeauthoryear{{F{\"o}rster Schreiber} \&
  {Wuyts}}{{F{\"o}rster Schreiber} \& {Wuyts}}{2020}]{forstershreiberwyuts20}
{F{\"o}rster Schreiber} N.~M.,  {Wuyts} S.,  2020, \mn@doi [\araa]
  {10.1146/annurev-astro-032620-021910}, \href
  {https://ui.adsabs.harvard.edu/abs/2020ARA&A..58..661F} {58, 661}

\bibitem[\protect\citeauthoryear{{Fudamoto} et~al.,}{{Fudamoto}
  et~al.}{2022}]{fudamoto22}
{Fudamoto} Y.,  et~al., 2022, \mn@doi [\apj] {10.3847/1538-4357/ac7a47}, \href
  {https://ui.adsabs.harvard.edu/abs/2022ApJ...934..144F} {934, 144}

\bibitem[\protect\citeauthoryear{{Gallagher} et~al.,}{{Gallagher}
  et~al.}{2018a}]{gallagher18a}
{Gallagher} M.~J.,  et~al., 2018a, \mn@doi [\apj] {10.3847/1538-4357/aabad8},
  \href {https://ui.adsabs.harvard.edu/abs/2018ApJ...858...90G} {858, 90}

\bibitem[\protect\citeauthoryear{{Gallagher} et~al.,}{{Gallagher}
  et~al.}{2018b}]{gallagher18b}
{Gallagher} M.~J.,  et~al., 2018b, \mn@doi [\apjl] {10.3847/2041-8213/aaf16a},
  \href {https://ui.adsabs.harvard.edu/abs/2018ApJ...868L..38G} {868, L38}

\bibitem[\protect\citeauthoryear{{Girma} \& {Teyssier}}{{Girma} \&
  {Teyssier}}{2024}]{girma24}
{Girma} E.,  {Teyssier} R.,  2024, \mn@doi [\mnras] {10.1093/mnras/stad3640},
  \href {https://ui.adsabs.harvard.edu/abs/2024MNRAS.527.6779G} {527, 6779}

\bibitem[\protect\citeauthoryear{{Grisdale}, {Agertz}, {Romeo}, {Renaud}  \&
  {Read}}{{Grisdale} et~al.}{2017}]{grisdale17}
{Grisdale} K.,  {Agertz} O.,  {Romeo} A.~B.,  {Renaud} F.,   {Read} J.~I.,
  2017, \mn@doi [\mnras] {10.1093/mnras/stw3133}, \href
  {https://ui.adsabs.harvard.edu/abs/2017MNRAS.466.1093G} {466, 1093}

\bibitem[\protect\citeauthoryear{{Grisdale}, {Agertz}, {Renaud}  \&
  {Romeo}}{{Grisdale} et~al.}{2018}]{grisdale18}
{Grisdale} K.,  {Agertz} O.,  {Renaud} F.,   {Romeo} A.~B.,  2018, \mn@doi
  [\mnras] {10.1093/mnras/sty1595}, \href
  {https://ui.adsabs.harvard.edu/abs/2018MNRAS.479.3167G} {479, 3167}

\bibitem[\protect\citeauthoryear{{Grisdale}, {Agertz}, {Renaud}, {Romeo},
  {Devriendt}  \& {Slyz}}{{Grisdale} et~al.}{2019}]{grisdale19}
{Grisdale} K.,  {Agertz} O.,  {Renaud} F.,  {Romeo} A.~B.,  {Devriendt} J.,
  {Slyz} A.,  2019, \mn@doi [\mnras] {10.1093/mnras/stz1201}, \href
  {https://ui.adsabs.harvard.edu/abs/2019MNRAS.486.5482G} {486, 5482}

\bibitem[\protect\citeauthoryear{{Grudi{\'c}}, {Hopkins},
  {Faucher-Gigu{\`e}re}, {Quataert}, {Murray}  \& {Kere{\v{s}}}}{{Grudi{\'c}}
  et~al.}{2018}]{grudic18}
{Grudi{\'c}} M.~Y.,  {Hopkins} P.~F.,  {Faucher-Gigu{\`e}re} C.-A.,  {Quataert}
  E.,  {Murray} N.,   {Kere{\v{s}}} D.,  2018, \mn@doi [\mnras]
  {10.1093/mnras/sty035}, \href
  {https://ui.adsabs.harvard.edu/abs/2018MNRAS.475.3511G} {475, 3511}

\bibitem[\protect\citeauthoryear{{Grudi{\'c}}, {Hopkins}, {Lee}, {Murray},
  {Faucher-Gigu{\`e}re}  \& {Johnson}}{{Grudi{\'c}} et~al.}{2019}]{grudic19}
{Grudi{\'c}} M.~Y.,  {Hopkins} P.~F.,  {Lee} E.~J.,  {Murray} N.,
  {Faucher-Gigu{\`e}re} C.-A.,   {Johnson} L.~C.,  2019, \mn@doi [\mnras]
  {10.1093/mnras/stz1758}, \href
  {https://ui.adsabs.harvard.edu/abs/2019MNRAS.488.1501G} {488, 1501}

\bibitem[\protect\citeauthoryear{{Haardt} \& {Madau}}{{Haardt} \&
  {Madau}}{1996}]{haardmadau96}
{Haardt} F.,  {Madau} P.,  1996, \mn@doi [\apj] {10.1086/177035}, \href
  {https://ui.adsabs.harvard.edu/abs/1996ApJ...461...20H} {461, 20}

\bibitem[\protect\citeauthoryear{{Hahn} \& {Abel}}{{Hahn} \&
  {Abel}}{2011}]{hannabel11}
{Hahn} O.,  {Abel} T.,  2011, \mn@doi [\mnras]
  {10.1111/j.1365-2966.2011.18820.x}, \href
  {https://ui.adsabs.harvard.edu/abs/2011MNRAS.415.2101H} {415, 2101}

\bibitem[\protect\citeauthoryear{{He}, {Bottrell}, {Wilson}, {Moreno},
  {Burkhart}, {Hayward}, {Hernquist}  \& {Twum}}{{He} et~al.}{2023}]{he23}
{He} H.,  {Bottrell} C.,  {Wilson} C.,  {Moreno} J.,  {Burkhart} B.,  {Hayward}
  C.~C.,  {Hernquist} L.,   {Twum} A.,  2023, \mn@doi [\apj]
  {10.3847/1538-4357/acca76}, \href
  {https://ui.adsabs.harvard.edu/abs/2023ApJ...950...56H} {950, 56}

\bibitem[\protect\citeauthoryear{{Heintz} et~al.,}{{Heintz}
  et~al.}{2022}]{heintz22}
{Heintz} K.~E.,  et~al., 2022, \mn@doi [\apjl] {10.3847/2041-8213/ac8057},
  \href {https://ui.adsabs.harvard.edu/abs/2022ApJ...934L..27H} {934, L27}

\bibitem[\protect\citeauthoryear{{Hennebelle} \& {Chabrier}}{{Hennebelle} \&
  {Chabrier}}{2011}]{hc11}
{Hennebelle} P.,  {Chabrier} G.,  2011, \mn@doi [\apjl]
  {10.1088/2041-8205/743/2/L29}, \href
  {https://ui.adsabs.harvard.edu/abs/2011ApJ...743L..29H} {743, L29}

\bibitem[\protect\citeauthoryear{{Herrera-Camus} et~al.,}{{Herrera-Camus}
  et~al.}{2022}]{herreracamus22}
{Herrera-Camus} R.,  et~al., 2022, \mn@doi [\aap]
  {10.1051/0004-6361/202142562}, \href
  {https://ui.adsabs.harvard.edu/abs/2022A&A...665L...8H} {665, L8}

\bibitem[\protect\citeauthoryear{{Heyer}, {Krawczyk}, {Duval}  \&
  {Jackson}}{{Heyer} et~al.}{2009}]{heyer09}
{Heyer} M.,  {Krawczyk} C.,  {Duval} J.,   {Jackson} J.~M.,  2009, \mn@doi
  [\apj] {10.1088/0004-637X/699/2/1092}, \href
  {https://ui.adsabs.harvard.edu/abs/2009ApJ...699.1092H} {699, 1092}

\bibitem[\protect\citeauthoryear{{Heyer}, {Gutermuth}, {Urquhart}, {Csengeri},
  {Wienen}, {Leurini}, {Menten}  \& {Wyrowski}}{{Heyer} et~al.}{2016}]{heyer16}
{Heyer} M.,  {Gutermuth} R.,  {Urquhart} J.~S.,  {Csengeri} T.,  {Wienen} M.,
  {Leurini} S.,  {Menten} K.,   {Wyrowski} F.,  2016, \mn@doi [\aap]
  {10.1051/0004-6361/201527681}, \href
  {https://ui.adsabs.harvard.edu/abs/2016A&A...588A..29H} {588, A29}

\bibitem[\protect\citeauthoryear{{Hopkins}}{{Hopkins}}{2013}]{hopkins13a}
{Hopkins} P.~F.,  2013, \mn@doi [\mnras] {10.1093/mnras/stt010}, \href
  {https://ui.adsabs.harvard.edu/abs/2013MNRAS.430.1880H} {430, 1880}

\bibitem[\protect\citeauthoryear{{Hopkins}, {Quataert}  \& {Murray}}{{Hopkins}
  et~al.}{2011}]{hopkins11}
{Hopkins} P.~F.,  {Quataert} E.,   {Murray} N.,  2011, \mn@doi [\mnras]
  {10.1111/j.1365-2966.2011.19306.x}, \href
  {https://ui.adsabs.harvard.edu/abs/2011MNRAS.417..950H} {417, 950}

\bibitem[\protect\citeauthoryear{{Hopkins}, {Quataert}  \& {Murray}}{{Hopkins}
  et~al.}{2012}]{hopkins12}
{Hopkins} P.~F.,  {Quataert} E.,   {Murray} N.,  2012, \mn@doi [\mnras]
  {10.1111/j.1365-2966.2012.20578.x}, \href
  {https://ui.adsabs.harvard.edu/abs/2012MNRAS.421.3488H} {421, 3488}

\bibitem[\protect\citeauthoryear{{Hopkins}, {Narayanan}  \& {Murray}}{{Hopkins}
  et~al.}{2013a}]{hopkins13c}
{Hopkins} P.~F.,  {Narayanan} D.,   {Murray} N.,  2013a, \mn@doi [\mnras]
  {10.1093/mnras/stt723}, \href
  {https://ui.adsabs.harvard.edu/abs/2013MNRAS.432.2647H} {432, 2647}

\bibitem[\protect\citeauthoryear{{Hopkins}, {Narayanan}, {Murray}  \&
  {Quataert}}{{Hopkins} et~al.}{2013b}]{hopkins13b}
{Hopkins} P.~F.,  {Narayanan} D.,  {Murray} N.,   {Quataert} E.,  2013b,
  \mn@doi [\mnras] {10.1093/mnras/stt688}, \href
  {https://ui.adsabs.harvard.edu/abs/2013MNRAS.433...69H} {433, 69}

\bibitem[\protect\citeauthoryear{{Hopkins}, {Kere{\v{s}}}, {O{\~n}orbe},
  {Faucher-Gigu{\`e}re}, {Quataert}, {Murray}  \& {Bullock}}{{Hopkins}
  et~al.}{2014}]{hopkins14}
{Hopkins} P.~F.,  {Kere{\v{s}}} D.,  {O{\~n}orbe} J.,  {Faucher-Gigu{\`e}re}
  C.-A.,  {Quataert} E.,  {Murray} N.,   {Bullock} J.~S.,  2014, \mn@doi
  [\mnras] {10.1093/mnras/stu1738}, \href
  {https://ui.adsabs.harvard.edu/abs/2014MNRAS.445..581H} {445, 581}

\bibitem[\protect\citeauthoryear{{Hopkins} et~al.,}{{Hopkins}
  et~al.}{2018}]{hopkins18}
{Hopkins} P.~F.,  et~al., 2018, \mn@doi [\mnras] {10.1093/mnras/sty1690}, \href
  {https://ui.adsabs.harvard.edu/abs/2018MNRAS.480..800H} {480, 800}

\bibitem[\protect\citeauthoryear{{Jim{\'e}nez-Donaire}
  et~al.,}{{Jim{\'e}nez-Donaire} et~al.}{2019}]{empire19}
{Jim{\'e}nez-Donaire} M.~J.,  et~al., 2019, \mn@doi [\apj]
  {10.3847/1538-4357/ab2b95}, \href
  {https://ui.adsabs.harvard.edu/abs/2019ApJ...880..127J} {880, 127}

\bibitem[\protect\citeauthoryear{{Kennicutt}}{{Kennicutt}}{1998}]{kennicutt98}
{Kennicutt} Robert~C. J.,  1998, \mn@doi [\apj] {10.1086/305588}, \href
  {https://ui.adsabs.harvard.edu/abs/1998ApJ...498..541K} {498, 541}

\bibitem[\protect\citeauthoryear{{Kim} \& {Ostriker}}{{Kim} \&
  {Ostriker}}{2015}]{kimostriker15}
{Kim} C.-G.,  {Ostriker} E.~C.,  2015, \mn@doi [\apj]
  {10.1088/0004-637X/802/2/99}, \href
  {https://ui.adsabs.harvard.edu/abs/2015ApJ...802...99K} {802, 99}

\bibitem[\protect\citeauthoryear{{Kim} et~al.,}{{Kim} et~al.}{2014}]{kim14}
{Kim} J.-h.,  et~al., 2014, \mn@doi [\apjs] {10.1088/0067-0049/210/1/14}, \href
  {https://ui.adsabs.harvard.edu/abs/2014ApJS..210...14K} {210, 14}

\bibitem[\protect\citeauthoryear{{Kim} et~al.,}{{Kim} et~al.}{2016}]{kim16}
{Kim} J.-h.,  et~al., 2016, \mn@doi [\apj] {10.3847/1538-4357/833/2/202}, \href
  {https://ui.adsabs.harvard.edu/abs/2016ApJ...833..202K} {833, 202}

\bibitem[\protect\citeauthoryear{{Klessen} \& {Glover}}{{Klessen} \&
  {Glover}}{2016}]{KlessenGlover2016}
{Klessen} R.~S.,  {Glover} S. C.~O.,  2016, \mn@doi [Saas-Fee Advanced Course]
  {10.1007/978-3-662-47890-5_2}, \href
  {https://ui.adsabs.harvard.edu/abs/2016SAAS...43...85K} {43, 85}

\bibitem[\protect\citeauthoryear{{Krahm}, {Finn}, {Indebetouw}, {Johnson},
  {Kamenetzky}  \& {Bemis}}{{Krahm} et~al.}{2024}]{krahm24}
{Krahm} G.,  {Finn} M.~K.,  {Indebetouw} R.,  {Johnson} K.~E.,  {Kamenetzky}
  J.,   {Bemis} A.,  2024, \mn@doi [\apj] {10.3847/1538-4357/ad2451}, \href
  {https://ui.adsabs.harvard.edu/abs/2024ApJ...964..166K} {964, 166}

\bibitem[\protect\citeauthoryear{{Kraljic}, {Renaud}, {Bournaud}, {Combes},
  {Elmegreen}, {Emsellem}  \& {Teyssier}}{{Kraljic} et~al.}{2014}]{kraljic14}
{Kraljic} K.,  {Renaud} F.,  {Bournaud} F.,  {Combes} F.,  {Elmegreen} B.,
  {Emsellem} E.,   {Teyssier} R.,  2014, \mn@doi [\apj]
  {10.1088/0004-637X/784/2/112}, \href
  {https://ui.adsabs.harvard.edu/abs/2014ApJ...784..112K} {784, 112}

\bibitem[\protect\citeauthoryear{{Kraljic} et~al.,}{{Kraljic}
  et~al.}{2024}]{kraljic24}
{Kraljic} K.,  et~al., 2024, \mn@doi [\aap] {10.1051/0004-6361/202347917},
  \href {https://ui.adsabs.harvard.edu/abs/2024A&A...682A..50K} {682, A50}

\bibitem[\protect\citeauthoryear{{Kretschmer} \& {Teyssier}}{{Kretschmer} \&
  {Teyssier}}{2020}]{kretschmer20}
{Kretschmer} M.,  {Teyssier} R.,  2020, \mn@doi [\mnras]
  {10.1093/mnras/stz3495}, \href
  {https://ui.adsabs.harvard.edu/abs/2020MNRAS.492.1385K} {492, 1385}

\bibitem[\protect\citeauthoryear{{Krumholz} \& {McKee}}{{Krumholz} \&
  {McKee}}{2005}]{km05}
{Krumholz} M.~R.,  {McKee} C.~F.,  2005, \mn@doi [\apj] {10.1086/431734}, \href
  {https://ui.adsabs.harvard.edu/abs/2005ApJ...630..250K} {630, 250}

\bibitem[\protect\citeauthoryear{{Krumholz}, {Dekel}  \& {McKee}}{{Krumholz}
  et~al.}{2012}]{krumholz12}
{Krumholz} M.~R.,  {Dekel} A.,   {McKee} C.~F.,  2012, \mn@doi [\apj]
  {10.1088/0004-637X/745/1/69}, \href
  {https://ui.adsabs.harvard.edu/abs/2012ApJ...745...69K} {745, 69}

\bibitem[\protect\citeauthoryear{{Krumholz}, {McKee}  \&
  {Bland-Hawthorn}}{{Krumholz} et~al.}{2019}]{Krumholz2019}
{Krumholz} M.~R.,  {McKee} C.~F.,   {Bland-Hawthorn} J.,  2019, \mn@doi [\araa]
  {10.1146/annurev-astro-091918-104430}, \href
  {https://ui.adsabs.harvard.edu/abs/2019ARA&A..57..227K} {57, 227}

\bibitem[\protect\citeauthoryear{{Larson}}{{Larson}}{1981}]{larson81}
{Larson} R.~B.,  1981, \mn@doi [\mnras] {10.1093/mnras/194.4.809}, \href
  {https://ui.adsabs.harvard.edu/abs/1981MNRAS.194..809L} {194, 809}

\bibitem[\protect\citeauthoryear{{Lee}, {Miville-Desch{\^e}nes}  \&
  {Murray}}{{Lee} et~al.}{2016}]{lee16}
{Lee} E.~J.,  {Miville-Desch{\^e}nes} M.-A.,   {Murray} N.~W.,  2016, \mn@doi
  [\apj] {10.3847/1538-4357/833/2/229}, \href
  {https://ui.adsabs.harvard.edu/abs/2016ApJ...833..229L} {833, 229}

\bibitem[\protect\citeauthoryear{{Leethochawalit} et~al.,}{{Leethochawalit}
  et~al.}{2023}]{leethochawalit23}
{Leethochawalit} N.,  et~al., 2023, \mn@doi [\apjl] {10.3847/2041-8213/ac959b},
  \href {https://ui.adsabs.harvard.edu/abs/2023ApJ...942L..26L} {942, L26}

\bibitem[\protect\citeauthoryear{{Leitherer} et~al.,}{{Leitherer}
  et~al.}{1999}]{leitherer99}
{Leitherer} C.,  et~al., 1999, \mn@doi [\apjs] {10.1086/313233}, \href
  {https://iopscience.iop.org/article/10.1086/313233} {123, 3}

\bibitem[\protect\citeauthoryear{{Leroy} et~al.,}{{Leroy}
  et~al.}{2013}]{leroy13}
{Leroy} A.~K.,  et~al., 2013, \mn@doi [\aj] {10.1088/0004-6256/146/2/19}, \href
  {https://ui.adsabs.harvard.edu/abs/2013AJ....146...19L} {146, 19}

\bibitem[\protect\citeauthoryear{{Leroy} et~al.,}{{Leroy}
  et~al.}{2015}]{leroy15}
{Leroy} A.~K.,  et~al., 2015, \mn@doi [\apj] {10.1088/0004-637X/801/1/25},
  \href {https://ui.adsabs.harvard.edu/abs/2015ApJ...801...25L} {801, 25}

\bibitem[\protect\citeauthoryear{{Leroy} et~al.,}{{Leroy}
  et~al.}{2016}]{leroy16}
{Leroy} A.~K.,  et~al., 2016, \mn@doi [\apj] {10.3847/0004-637X/831/1/16},
  \href {https://ui.adsabs.harvard.edu/abs/2016ApJ...831...16L} {831, 16}

\bibitem[\protect\citeauthoryear{{Leroy} et~al.,}{{Leroy}
  et~al.}{2017}]{leroy17}
{Leroy} A.~K.,  et~al., 2017, \mn@doi [\apj] {10.3847/1538-4357/aa7fef}, \href
  {https://ui.adsabs.harvard.edu/abs/2017ApJ...846...71L} {846, 71}

\bibitem[\protect\citeauthoryear{{Li}, {Gnedin}  \& {Gnedin}}{{Li}
  et~al.}{2018}]{li19}
{Li} H.,  {Gnedin} O.~Y.,   {Gnedin} N.~Y.,  2018, \mn@doi [\apj]
  {10.3847/1538-4357/aac9b8}, \href
  {https://ui.adsabs.harvard.edu/abs/2018ApJ...861..107L} {861, 107}

\bibitem[\protect\citeauthoryear{{Li}, {Vogelsberger}, {Marinacci}, {Sales}  \&
  {Torrey}}{{Li} et~al.}{2020}]{li20}
{Li} H.,  {Vogelsberger} M.,  {Marinacci} F.,  {Sales} L.~V.,   {Torrey} P.,
  2020, \mn@doi [\mnras] {10.1093/mnras/staa3122}, \href
  {https://ui.adsabs.harvard.edu/abs/2020MNRAS.499.5862L} {499, 5862}

\bibitem[\protect\citeauthoryear{{Li}, {Vogelsberger}, {Bryan}, {Marinacci},
  {Sales}  \& {Torrey}}{{Li} et~al.}{2022}]{li22}
{Li} H.,  {Vogelsberger} M.,  {Bryan} G.~L.,  {Marinacci} F.,  {Sales} L.~V.,
  {Torrey} P.,  2022, \mn@doi [\mnras] {10.1093/mnras/stac1136}, \href
  {https://ui.adsabs.harvard.edu/abs/2022MNRAS.514..265L} {514, 265}

\bibitem[\protect\citeauthoryear{{Mac Low}}{{Mac Low}}{1999}]{maclow99}
{Mac Low} M.-M.,  1999, \mn@doi [\apj] {10.1086/307784}, \href
  {https://ui.adsabs.harvard.edu/abs/1999ApJ...524..169M} {524, 169}

\bibitem[\protect\citeauthoryear{{Mac Low}, {Klessen}, {Burkert}  \&
  {Smith}}{{Mac Low} et~al.}{1998}]{maclow98}
{Mac Low} M.-M.,  {Klessen} R.~S.,  {Burkert} A.,   {Smith} M.~D.,  1998,
  \mn@doi [\prl] {10.1103/PhysRevLett.80.2754}, \href
  {https://ui.adsabs.harvard.edu/abs/1998PhRvL..80.2754M} {80, 2754}

\bibitem[\protect\citeauthoryear{{McKee} \& {Ostriker}}{{McKee} \&
  {Ostriker}}{2007}]{McKeeOstriker2007}
{McKee} C.~F.,  {Ostriker} E.~C.,  2007, \mn@doi [\araa]
  {10.1146/annurev.astro.45.051806.110602}, \href
  {https://ui.adsabs.harvard.edu/abs/2007ARA&A..45..565M} {45, 565}

\bibitem[\protect\citeauthoryear{{Miville-Desch{\^e}nes}, {Murray}  \&
  {Lee}}{{Miville-Desch{\^e}nes} et~al.}{2017}]{mivilledeschenes17}
{Miville-Desch{\^e}nes} M.-A.,  {Murray} N.,   {Lee} E.~J.,  2017, \mn@doi
  [\apj] {10.3847/1538-4357/834/1/57}, \href
  {https://ui.adsabs.harvard.edu/abs/2017ApJ...834...57M} {834, 57}

\bibitem[\protect\citeauthoryear{M{\"u}ller}{M{\"u}ller}{2000}]{muller00}
M{\"u}ller J.,  2000, \mn@doi [Journal of Research of the National Institute of
  Standards and Technology] {10.6028/jres.105.044}, 105, 551

\bibitem[\protect\citeauthoryear{{Neumann} et~al.,}{{Neumann}
  et~al.}{2023}]{neumann23}
{Neumann} L.,  et~al., 2023, \mn@doi [\mnras] {10.1093/mnras/stad424}, \href
  {https://ui.adsabs.harvard.edu/abs/2023MNRAS.521.3348N} {521, 3348}

\bibitem[\protect\citeauthoryear{{Nobels}, {Schaye}, {Schaller}, {Ploeckinger},
  {Chaikin}  \& {Richings}}{{Nobels} et~al.}{2024}]{nobles24}
{Nobels} F. S.~J.,  {Schaye} J.,  {Schaller} M.,  {Ploeckinger} S.,  {Chaikin}
  E.,   {Richings} A.~J.,  2024, \mn@doi [\mnras] {10.1093/mnras/stae1390},
  \href {https://ui.adsabs.harvard.edu/abs/2024MNRAS.532.3299N} {532, 3299}

\bibitem[\protect\citeauthoryear{{Nu{\~n}ez-Casti{\~n}eyra}, {Nezri},
  {Devriendt}  \& {Teyssier}}{{Nu{\~n}ez-Casti{\~n}eyra}
  et~al.}{2021}]{arturo21}
{Nu{\~n}ez-Casti{\~n}eyra} A.,  {Nezri} E.,  {Devriendt} J.,   {Teyssier} R.,
  2021, \mn@doi [\mnras] {10.1093/mnras/staa3233}, \href
  {https://ui.adsabs.harvard.edu/abs/2021MNRAS.501...62N} {501, 62}

\bibitem[\protect\citeauthoryear{{Ohlin}, {Renaud}  \& {Agertz}}{{Ohlin}
  et~al.}{2019}]{ohlin19}
{Ohlin} L.,  {Renaud} F.,   {Agertz} O.,  2019, \mn@doi [\mnras]
  {10.1093/mnras/stz705}, \href
  {https://ui.adsabs.harvard.edu/abs/2019MNRAS.485.3887O} {485, 3887}

\bibitem[\protect\citeauthoryear{{Orr} et~al.,}{{Orr} et~al.}{2018}]{orr18}
{Orr} M.~E.,  et~al., 2018, \mn@doi [\mnras] {10.1093/mnras/sty1241}, \href
  {https://ui.adsabs.harvard.edu/abs/2018MNRAS.478.3653O} {478, 3653}

\bibitem[\protect\citeauthoryear{{Padoan} \& {Nordlund}}{{Padoan} \&
  {Nordlund}}{2011}]{pn11}
{Padoan} P.,  {Nordlund} {\r{A}}.,  2011, \mn@doi [\apj]
  {10.1088/0004-637X/730/1/40}, \href
  {https://ui.adsabs.harvard.edu/abs/2011ApJ...730...40P} {730, 40}

\bibitem[\protect\citeauthoryear{{Padoan}, {Haugb{\o}lle}  \&
  {Nordlund}}{{Padoan} et~al.}{2012}]{pn12}
{Padoan} P.,  {Haugb{\o}lle} T.,   {Nordlund} {\r{A}}.,  2012, \mn@doi [\apjl]
  {10.1088/2041-8205/759/2/L27}, \href
  {https://ui.adsabs.harvard.edu/abs/2012ApJ...759L..27P} {759, L27}

\bibitem[\protect\citeauthoryear{{Parlanti}, {Carniani}, {Pallottini},
  {Cignoni}, {Cresci}, {Kohandel}, {Mannucci}  \& {Marconi}}{{Parlanti}
  et~al.}{2023}]{parlanti23}
{Parlanti} E.,  {Carniani} S.,  {Pallottini} A.,  {Cignoni} M.,  {Cresci} G.,
  {Kohandel} M.,  {Mannucci} F.,   {Marconi} A.,  2023, \mn@doi [\aap]
  {10.1051/0004-6361/202245603}, \href
  {https://ui.adsabs.harvard.edu/abs/2023A&A...673A.153P} {673, A153}

\bibitem[\protect\citeauthoryear{{Petkova} et~al.,}{{Petkova}
  et~al.}{2023}]{petkova23}
{Petkova} M.~A.,  et~al., 2023, \mn@doi [\mnras] {10.1093/mnras/stad2344},
  \href {https://ui.adsabs.harvard.edu/abs/2023MNRAS.525..962P} {525, 962}

\bibitem[\protect\citeauthoryear{{Polzin}, {Kravtsov}, {Semenov}  \&
  {Gnedin}}{{Polzin} et~al.}{2024}]{polzin24}
{Polzin} A.,  {Kravtsov} A.~V.,  {Semenov} V.~A.,   {Gnedin} N.~Y.,  2024,
  \mn@doi [arXiv e-prints] {10.48550/arXiv.2407.11125}, \href
  {https://ui.adsabs.harvard.edu/abs/2024arXiv240711125P} {p. arXiv:2407.11125}

\bibitem[\protect\citeauthoryear{{Pope} et~al.,}{{Pope} et~al.}{2023}]{pope23}
{Pope} A.,  et~al., 2023, \mn@doi [\apjl] {10.3847/2041-8213/acdf5a}, \href
  {https://ui.adsabs.harvard.edu/abs/2023ApJ...951L..46P} {951, L46}

\bibitem[\protect\citeauthoryear{{Posses} et~al.,}{{Posses}
  et~al.}{2023}]{posses23}
{Posses} A.~C.,  et~al., 2023, \mn@doi [\aap] {10.1051/0004-6361/202243399},
  \href {https://ui.adsabs.harvard.edu/abs/2023A&A...669A..46P} {669, A46}

\bibitem[\protect\citeauthoryear{{Querejeta} et~al.,}{{Querejeta}
  et~al.}{2019}]{querejeta19}
{Querejeta} M.,  et~al., 2019, \mn@doi [\aap] {10.1051/0004-6361/201834915},
  \href {https://ui.adsabs.harvard.edu/abs/2019A&A...625A..19Q} {625, A19}

\bibitem[\protect\citeauthoryear{{Raiteri}, {Villata}  \& {Navarro}}{{Raiteri}
  et~al.}{1996}]{raiteri96}
{Raiteri} C.~M.,  {Villata} M.,   {Navarro} J.~F.,  1996, \aap, \href
  {https://ui.adsabs.harvard.edu/abs/1996A&A...315..105R} {315, 105}

\bibitem[\protect\citeauthoryear{{Renaud}, {Kraljic}  \& {Bournaud}}{{Renaud}
  et~al.}{2012}]{renaud12}
{Renaud} F.,  {Kraljic} K.,   {Bournaud} F.,  2012, \mn@doi [\apjl]
  {10.1088/2041-8205/760/1/L16}, \href
  {https://ui.adsabs.harvard.edu/abs/2012ApJ...760L..16R} {760, L16}

\bibitem[\protect\citeauthoryear{{Renaud} et~al.,}{{Renaud}
  et~al.}{2013}]{renaud13}
{Renaud} F.,  et~al., 2013, \mn@doi [\mnras] {10.1093/mnras/stt1698}, \href
  {https://ui.adsabs.harvard.edu/abs/2013MNRAS.436.1836R} {436, 1836}

\bibitem[\protect\citeauthoryear{{Renaud}, {Bournaud}, {Kraljic}  \&
  {Duc}}{{Renaud} et~al.}{2014}]{renaud14}
{Renaud} F.,  {Bournaud} F.,  {Kraljic} K.,   {Duc} P.~A.,  2014, \mn@doi
  [\mnras] {10.1093/mnrasl/slu050}, \href
  {https://ui.adsabs.harvard.edu/abs/2014MNRAS.442L..33R} {442, L33}

\bibitem[\protect\citeauthoryear{{Renaud}, {Agertz}, {Read}, {Ryde},
  {Andersson}, {Bensby}, {Rey}  \& {Feuillet}}{{Renaud}
  et~al.}{2021a}]{vinter2}
{Renaud} F.,  {Agertz} O.,  {Read} J.~I.,  {Ryde} N.,  {Andersson} E.~P.,
  {Bensby} T.,  {Rey} M.~P.,   {Feuillet} D.~K.,  2021a, \mn@doi [\mnras]
  {10.1093/mnras/stab250}, \href
  {https://ui.adsabs.harvard.edu/abs/2021MNRAS.503.5846R} {503, 5846}

\bibitem[\protect\citeauthoryear{{Renaud}, {Agertz}, {Andersson}, {Read},
  {Ryde}, {Bensby}, {Rey}  \& {Feuillet}}{{Renaud} et~al.}{2021b}]{vinter3}
{Renaud} F.,  {Agertz} O.,  {Andersson} E.~P.,  {Read} J.~I.,  {Ryde} N.,
  {Bensby} T.,  {Rey} M.~P.,   {Feuillet} D.~K.,  2021b, \mn@doi [\mnras]
  {10.1093/mnras/stab543}, \href
  {https://ui.adsabs.harvard.edu/abs/2021MNRAS.503.5868R} {503, 5868}

\bibitem[\protect\citeauthoryear{{Renaud}, {Segovia Otero}  \&
  {Agertz}}{{Renaud} et~al.}{2022}]{renaud22}
{Renaud} F.,  {Segovia Otero} {\'A}.,   {Agertz} O.,  2022, \mn@doi [\mnras]
  {10.1093/mnras/stac2557}, \href
  {https://ui.adsabs.harvard.edu/abs/2022MNRAS.516.4922R} {516, 4922}

\bibitem[\protect\citeauthoryear{{Renaud}, {Agertz}  \& {Romeo}}{{Renaud}
  et~al.}{2024}]{renaud24}
{Renaud} F.,  {Agertz} O.,   {Romeo} A.~B.,  2024, \mn@doi [\aap]
  {10.1051/0004-6361/202449721}, \href
  {https://ui.adsabs.harvard.edu/abs/2024A&A...687A..91R} {687, A91}

\bibitem[\protect\citeauthoryear{{Rizzo} et~al.,}{{Rizzo}
  et~al.}{2023}]{rizzo23}
{Rizzo} F.,  et~al., 2023, \mn@doi [arXiv e-prints]
  {10.48550/arXiv.2303.16227}, \href
  {https://ui.adsabs.harvard.edu/abs/2023arXiv230316227R} {p. arXiv:2303.16227}

\bibitem[\protect\citeauthoryear{{Roca-F{\`a}brega} et~al.,}{{Roca-F{\`a}brega}
  et~al.}{2021}]{santi21}
{Roca-F{\`a}brega} S.,  et~al., 2021, \mn@doi [\apj]
  {10.3847/1538-4357/ac088a}, \href
  {https://ui.adsabs.harvard.edu/abs/2021ApJ...917...64R} {917, 64}

\bibitem[\protect\citeauthoryear{{Rodighiero} et~al.,}{{Rodighiero}
  et~al.}{2011}]{rodighiero11}
{Rodighiero} G.,  et~al., 2011, \mn@doi [\apjl] {10.1088/2041-8205/739/2/L40},
  \href {https://ui.adsabs.harvard.edu/abs/2011ApJ...739L..40R} {739, L40}

\bibitem[\protect\citeauthoryear{{Roman-Oliveira}, {Fraternali}  \&
  {Rizzo}}{{Roman-Oliveira} et~al.}{2023}]{romanolveira23}
{Roman-Oliveira} F.,  {Fraternali} F.,   {Rizzo} F.,  2023, \mn@doi [\mnras]
  {10.1093/mnras/stad530}, \href
  {https://ui.adsabs.harvard.edu/abs/2023MNRAS.521.1045R} {521, 1045}

\bibitem[\protect\citeauthoryear{{Romeo}, {Agertz}  \& {Renaud}}{{Romeo}
  et~al.}{2023}]{romeo23}
{Romeo} A.~B.,  {Agertz} O.,   {Renaud} F.,  2023, \mn@doi [\mnras]
  {10.1093/mnras/stac3074}, \href
  {https://ui.adsabs.harvard.edu/abs/2023MNRAS.518.1002R} {518, 1002}

\bibitem[\protect\citeauthoryear{{Rosen} \& {Bregman}}{{Rosen} \&
  {Bregman}}{1995}]{rosenbergmas95}
{Rosen} A.,  {Bregman} J.~N.,  1995, \mn@doi [\apj] {10.1086/175303}, \href
  {https://ui.adsabs.harvard.edu/abs/1995ApJ...440..634R} {440, 634}

\bibitem[\protect\citeauthoryear{{Rosolowsky} \& {Leroy}}{{Rosolowsky} \&
  {Leroy}}{2006}]{rosolowskyleroy06}
{Rosolowsky} E.,  {Leroy} A.,  2006, \mn@doi [\pasp] {10.1086/502982}, \href
  {https://ui.adsabs.harvard.edu/abs/2006PASP..118..590R} {118, 590}

\bibitem[\protect\citeauthoryear{{Rosolowsky} et~al.,}{{Rosolowsky}
  et~al.}{2021}]{rosolowsky21}
{Rosolowsky} E.,  et~al., 2021, \mn@doi [\mnras] {10.1093/mnras/stab085}, \href
  {https://ui.adsabs.harvard.edu/abs/2021MNRAS.502.1218R} {502, 1218}

\bibitem[\protect\citeauthoryear{{Saintonge} \& {Catinella}}{{Saintonge} \&
  {Catinella}}{2022}]{saintonge22}
{Saintonge} A.,  {Catinella} B.,  2022, \mn@doi [\araa]
  {10.1146/annurev-astro-021022-043545}, \href
  {https://ui.adsabs.harvard.edu/abs/2022ARA&A..60..319S} {60, 319}

\bibitem[\protect\citeauthoryear{{Schinnerer} \& {Leroy}}{{Schinnerer} \&
  {Leroy}}{2024}]{schinnererleroy24}
{Schinnerer} E.,  {Leroy} A.~K.,  2024, \mn@doi [arXiv e-prints]
  {10.48550/arXiv.2403.19843}, \href
  {https://ui.adsabs.harvard.edu/abs/2024arXiv240319843S} {p. arXiv:2403.19843}

\bibitem[\protect\citeauthoryear{{Schmidt} \& {Federrath}}{{Schmidt} \&
  {Federrath}}{2011}]{schmidtfederrath11}
{Schmidt} W.,  {Federrath} C.,  2011, \mn@doi [\aap]
  {10.1051/0004-6361/201015630}, \href
  {https://ui.adsabs.harvard.edu/abs/2011A&A...528A.106S} {528, A106}

\bibitem[\protect\citeauthoryear{{Schmidt} et~al.,}{{Schmidt}
  et~al.}{2014}]{schmidt14}
{Schmidt} W.,  et~al., 2014, \mn@doi [\mnras] {10.1093/mnras/stu501}, \href
  {https://ui.adsabs.harvard.edu/abs/2014MNRAS.440.3051S} {440, 3051}

\bibitem[\protect\citeauthoryear{{Scholte} et~al.,}{{Scholte}
  et~al.}{2024}]{Scholte2024}
{Scholte} D.,  et~al., 2024, \mn@doi [arXiv e-prints]
  {10.48550/arXiv.2408.03996}, \href
  {https://ui.adsabs.harvard.edu/abs/2024arXiv240803996S} {p. arXiv:2408.03996}

\bibitem[\protect\citeauthoryear{{Schruba}, {Kruijssen}  \& {Leroy}}{{Schruba}
  et~al.}{2019}]{schruba19}
{Schruba} A.,  {Kruijssen} J.~M.~D.,   {Leroy} A.~K.,  2019, \mn@doi [\apj]
  {10.3847/1538-4357/ab3a43}, \href
  {https://ui.adsabs.harvard.edu/abs/2019ApJ...883....2S} {883, 2}

\bibitem[\protect\citeauthoryear{{Segovia Otero}, {Renaud}  \&
  {Agertz}}{{Segovia Otero} et~al.}{2022}]{aso22}
{Segovia Otero} {\'A}.,  {Renaud} F.,   {Agertz} O.,  2022, \mn@doi [\mnras]
  {10.1093/mnras/stac2368}, \href
  {https://ui.adsabs.harvard.edu/abs/2022MNRAS.516.2272S} {516, 2272}

\bibitem[\protect\citeauthoryear{{Semenov}, {Kravtsov}  \& {Gnedin}}{{Semenov}
  et~al.}{2016}]{semenov16}
{Semenov} V.~A.,  {Kravtsov} A.~V.,   {Gnedin} N.~Y.,  2016, \mn@doi [\apj]
  {10.3847/0004-637X/826/2/200}, \href
  {https://ui.adsabs.harvard.edu/abs/2016ApJ...826..200S} {826, 200}

\bibitem[\protect\citeauthoryear{{Semenov}, {Kravtsov}  \& {Gnedin}}{{Semenov}
  et~al.}{2017}]{semenov17}
{Semenov} V.~A.,  {Kravtsov} A.~V.,   {Gnedin} N.~Y.,  2017, \mn@doi [\apj]
  {10.3847/1538-4357/aa8096}, \href
  {https://ui.adsabs.harvard.edu/abs/2017ApJ...845..133S} {845, 133}

\bibitem[\protect\citeauthoryear{{Semenov}, {Kravtsov}  \& {Gnedin}}{{Semenov}
  et~al.}{2018}]{semenov18}
{Semenov} V.~A.,  {Kravtsov} A.~V.,   {Gnedin} N.~Y.,  2018, \mn@doi [\apj]
  {10.3847/1538-4357/aac6eb}, \href
  {https://ui.adsabs.harvard.edu/abs/2018ApJ...861....4S} {861, 4}

\bibitem[\protect\citeauthoryear{{Solomon}, {Rivolo}, {Barrett}  \&
  {Yahil}}{{Solomon} et~al.}{1987}]{Solomon1987}
{Solomon} P.~M.,  {Rivolo} A.~R.,  {Barrett} J.,   {Yahil} A.,  1987, \mn@doi
  [\apj] {10.1086/165493}, \href
  {https://ui.adsabs.harvard.edu/abs/1987ApJ...319..730S} {319, 730}

\bibitem[\protect\citeauthoryear{{Speagle}, {Steinhardt}, {Capak}  \&
  {Silverman}}{{Speagle} et~al.}{2014}]{speagle14}
{Speagle} J.~S.,  {Steinhardt} C.~L.,  {Capak} P.~L.,   {Silverman} J.~D.,
  2014, \mn@doi [\apjs] {10.1088/0067-0049/214/2/15}, \href
  {https://ui.adsabs.harvard.edu/abs/2014ApJS..214...15S} {214, 15}

\bibitem[\protect\citeauthoryear{{Stone}, {Ostriker}  \& {Gammie}}{{Stone}
  et~al.}{1998}]{Stone1998}
{Stone} J.~M.,  {Ostriker} E.~C.,   {Gammie} C.~F.,  1998, \mn@doi [\apjl]
  {10.1086/311718}, \href
  {https://ui.adsabs.harvard.edu/abs/1998ApJ...508L..99S} {508, L99}

\bibitem[\protect\citeauthoryear{{Sun} et~al.,}{{Sun} et~al.}{2018}]{sun18}
{Sun} J.,  et~al., 2018, \mn@doi [\apj] {10.3847/1538-4357/aac326}, \href
  {https://ui.adsabs.harvard.edu/abs/2018ApJ...860..172S} {860, 172}

\bibitem[\protect\citeauthoryear{{Sun} et~al.,}{{Sun} et~al.}{2020a}]{sun20a}
{Sun} J.,  et~al., 2020a, \mn@doi [\apj] {10.3847/1538-4357/ab781c}, \href
  {https://ui.adsabs.harvard.edu/abs/2020ApJ...892..148S} {892, 148}

\bibitem[\protect\citeauthoryear{{Sun} et~al.,}{{Sun} et~al.}{2020b}]{sun20b}
{Sun} J.,  et~al., 2020b, \mn@doi [\apjl] {10.3847/2041-8213/abb3be}, \href
  {https://ui.adsabs.harvard.edu/abs/2020ApJ...901L...8S} {901, L8}

\bibitem[\protect\citeauthoryear{{Sun} et~al.,}{{Sun} et~al.}{2022}]{sun22}
{Sun} J.,  et~al., 2022, \mn@doi [\aj] {10.3847/1538-3881/ac74bd}, \href
  {https://ui.adsabs.harvard.edu/abs/2022AJ....164...43S} {164, 43}

\bibitem[\protect\citeauthoryear{{Sun} et~al.,}{{Sun} et~al.}{2023}]{sun23}
{Sun} J.,  et~al., 2023, \mn@doi [\apjl] {10.3847/2041-8213/acbd9c}, \href
  {https://ui.adsabs.harvard.edu/abs/2023ApJ...945L..19S} {945, L19}

\bibitem[\protect\citeauthoryear{{Sutherland} \& {Dopita}}{{Sutherland} \&
  {Dopita}}{1993}]{suthdopita93}
{Sutherland} R.~S.,  {Dopita} M.~A.,  1993, \mn@doi [\apjs] {10.1086/191823},
  \href {https://ui.adsabs.harvard.edu/abs/1993ApJS...88..253S} {88, 253}

\bibitem[\protect\citeauthoryear{{Tacconi} et~al.,}{{Tacconi}
  et~al.}{2018}]{tacconi18}
{Tacconi} L.~J.,  et~al., 2018, \mn@doi [\apj] {10.3847/1538-4357/aaa4b4},
  \href {https://ui.adsabs.harvard.edu/abs/2018ApJ...853..179T} {853, 179}

\bibitem[\protect\citeauthoryear{{Tacconi}, {Genzel}  \& {Sternberg}}{{Tacconi}
  et~al.}{2020}]{tacconi20}
{Tacconi} L.~J.,  {Genzel} R.,   {Sternberg} A.,  2020, \mn@doi [\araa]
  {10.1146/annurev-astro-082812-141034}, \href
  {https://ui.adsabs.harvard.edu/abs/2020ARA&A..58..157T} {58, 157}

\bibitem[\protect\citeauthoryear{{Tamburro}, {Rix}, {Leroy}, {Mac Low},
  {Walter}, {Kennicutt}, {Brinks}  \& {de Blok}}{{Tamburro}
  et~al.}{2009}]{tamburro09}
{Tamburro} D.,  {Rix} H.~W.,  {Leroy} A.~K.,  {Mac Low} M.~M.,  {Walter} F.,
  {Kennicutt} R.~C.,  {Brinks} E.,   {de Blok} W.~J.~G.,  2009, \mn@doi [\aj]
  {10.1088/0004-6256/137/5/4424}, \href
  {https://ui.adsabs.harvard.edu/abs/2009AJ....137.4424T} {137, 4424}

\bibitem[\protect\citeauthoryear{{Teyssier}}{{Teyssier}}{2002}]{teyssier02}
{Teyssier} R.,  2002, \mn@doi [\aap] {10.1051/0004-6361:20011817}, \href
  {https://ui.adsabs.harvard.edu/abs/2002A&A...385..337T} {385, 337}

\bibitem[\protect\citeauthoryear{{Trebitsch}, {Blaizot}, {Rosdahl}, {Devriendt}
   \& {Slyz}}{{Trebitsch} et~al.}{2017}]{trebitsch17}
{Trebitsch} M.,  {Blaizot} J.,  {Rosdahl} J.,  {Devriendt} J.,   {Slyz} A.,
  2017, \mn@doi [\mnras] {10.1093/mnras/stx1060}, \href
  {https://ui.adsabs.harvard.edu/abs/2017MNRAS.470..224T} {470, 224}

\bibitem[\protect\citeauthoryear{{Trussler} et~al.,}{{Trussler}
  et~al.}{2023}]{trussler23}
{Trussler} J. A.~A.,  et~al., 2023, \mn@doi [\mnras] {10.1093/mnras/stad1629},
  \href {https://ui.adsabs.harvard.edu/abs/2023MNRAS.523.3423T} {523, 3423}

\bibitem[\protect\citeauthoryear{{{\"U}bler} et~al.,}{{{\"U}bler}
  et~al.}{2019}]{ubler19}
{{\"U}bler} H.,  et~al., 2019, \mn@doi [\apj] {10.3847/1538-4357/ab27cc}, \href
  {https://ui.adsabs.harvard.edu/abs/2019ApJ...880...48U} {880, 48}

\bibitem[\protect\citeauthoryear{{Utomo} et~al.,}{{Utomo}
  et~al.}{2018}]{utomo18}
{Utomo} D.,  et~al., 2018, \mn@doi [\apjl] {10.3847/2041-8213/aacf8f}, \href
  {https://ui.adsabs.harvard.edu/abs/2018ApJ...861L..18U} {861, L18}

\bibitem[\protect\citeauthoryear{{Wei}, {Keto}  \& {Ho}}{{Wei}
  et~al.}{2012}]{wei12}
{Wei} L.~H.,  {Keto} E.,   {Ho} L.~C.,  2012, \mn@doi [\apj]
  {10.1088/0004-637X/750/2/136}, \href
  {https://ui.adsabs.harvard.edu/abs/2012ApJ...750..136W} {750, 136}

\bibitem[\protect\citeauthoryear{{Wetzel} et~al.,}{{Wetzel}
  et~al.}{2023}]{Wetzel2023}
{Wetzel} A.,  et~al., 2023, \mn@doi [\apjs] {10.3847/1538-4365/acb99a}, \href
  {https://ui.adsabs.harvard.edu/abs/2023ApJS..265...44W} {265, 44}

\bibitem[\protect\citeauthoryear{{Wilson}, {Bemis}, {Ledger}  \&
  {Klimi}}{{Wilson} et~al.}{2023}]{wilson23}
{Wilson} C.~D.,  {Bemis} A.,  {Ledger} B.,   {Klimi} O.,  2023, \mn@doi
  [\mnras] {10.1093/mnras/stad560}, \href
  {https://ui.adsabs.harvard.edu/abs/2023MNRAS.521..717W} {521, 717}

\bibitem[\protect\citeauthoryear{{Wise}, {Turk}, {Norman}  \& {Abel}}{{Wise}
  et~al.}{2012}]{wise12}
{Wise} J.~H.,  {Turk} M.~J.,  {Norman} M.~L.,   {Abel} T.,  2012, \mn@doi
  [\apj] {10.1088/0004-637X/745/1/50}, \href
  {https://ui.adsabs.harvard.edu/abs/2012ApJ...745...50W} {745, 50}

\bibitem[\protect\citeauthoryear{{Wisnioski} et~al.,}{{Wisnioski}
  et~al.}{2015}]{wisnioski15}
{Wisnioski} E.,  et~al., 2015, \mn@doi [\apj] {10.1088/0004-637X/799/2/209},
  \href {https://ui.adsabs.harvard.edu/abs/2015ApJ...799..209W} {799, 209}

\bibitem[\protect\citeauthoryear{{Woosley} \& {Heger}}{{Woosley} \&
  {Heger}}{2007}]{woosleyheger07}
{Woosley} S.~E.,  {Heger} A.,  2007, \mn@doi [\physrep]
  {10.1016/j.physrep.2007.02.009}, \href
  {https://ui.adsabs.harvard.edu/abs/2007PhR...442..269W} {442, 269}

\bibitem[\protect\citeauthoryear{{Wu}, {Evans}, {Shirley}  \& {Knez}}{{Wu}
  et~al.}{2010}]{wu10}
{Wu} J.,  {Evans} Neal~J. I.,  {Shirley} Y.~L.,   {Knez} C.,  2010, \mn@doi
  [\apjs] {10.1088/0067-0049/188/2/313}, \href
  {https://ui.adsabs.harvard.edu/abs/2010ApJS..188..313W} {188, 313}

\bibitem[\protect\citeauthoryear{{Wuyts} et~al.,}{{Wuyts}
  et~al.}{2011}]{wuyts11}
{Wuyts} S.,  et~al., 2011, \mn@doi [\apj] {10.1088/0004-637X/742/2/96}, \href
  {https://ui.adsabs.harvard.edu/abs/2011ApJ...742...96W} {742, 96}

\bibitem[\protect\citeauthoryear{{Zakardjian} et~al.,}{{Zakardjian}
  et~al.}{2023}]{zakardjian23}
{Zakardjian} A.,  et~al., 2023, \mn@doi [\aap] {10.1051/0004-6361/202244520},
  \href {https://ui.adsabs.harvard.edu/abs/2023A&A...678A.171Z} {678, A171}

\makeatother
\end{thebibliography}

%
%
%
%
%
%
%
%

%

%

%

%
\bsp	%
\label{lastpage}
\end{document}